\documentclass[showkeys,aps,showpacs,nofootinbib]{revtex4-1}

\usepackage{amsmath}
\usepackage{amsfonts}
\usepackage{amssymb,amscd,amsbsy}
\usepackage{epsfig}
\usepackage{color}
\usepackage{threeparttable}  

\usepackage{mathtools}

\begin{document}

\title{Test MaxEnt in Social Strategy Transitions with Experimental Two-Person Constant Sum 2$\times$2 Games}

\author{Bin Xu$^{1,2}$}
\author{Zhijian Wang$^{2}$\footnote{Corresponding author: Tel.: +86 13905815529; fax: +86 571 87176578. \\E-mail addresses: wangzj@zju.edu.cn (Z.J. Wang)}}

\address{$^{1}$Public Administration College, Zhejiang Gongshang University, Hangzhou, 310018, China}
\address{$^{2}$Experimental Social Science Laboratory, Zhejiang University, Hangzhou, 310058, China}


\begin{abstract}

\begin{center}
Abstract
\end{center}

Using laboratory experimental data, we test the uncertainty of social state transitions in various competing environments of fixed paired two-person constant sum $2 \times 2$ games. It firstly shows that, the distributions of social strategy transitions are not erratic but obey the principle of the maximum entropy (MaxEnt). This finding indicates that human subject social systems and natural systems could share wider common backgrounds.
\end{abstract}



\maketitle

\section{Introduction}
The principle of the maximum entropy (MaxEnt) is introduced by Jaynes~\cite{Jaynes1957}, rooting in Boltzmann, Gibbs and Shannon~\cite{Shannon1948,Jaynes2003}.
As a methodology, MaxEnt has gained its wide applications in natural science and engineering. The advantage of this methodology is to provide rich information based on very limit information. In economics, MaxEnt approach has also gained its wide applications, e.g., in  market equilibrium~\cite{Toda2010,Barde2012}, in wealth and income distribution~\cite{Castaldi2007,WuMaxEntIncome2003},  in firm growth rates~\cite{Alfarano2008} and in behavior modeling~\cite{Wolpert2012}. Theoretical interpreting or modeling of the distributions of social outcomes with MaxEnt is growing.

Considering the importance of MaxEnt, to carry out laboratory experiments to investigate this fundamental rule is necessary~\cite{Falk2009}. Only quite recently, entropy is firstly measured in experimental economics systems to evaluate social outcomes by Bednar et.al.~\cite{Yan2011} and Cason et.al~\cite{cason2011behavioral}. Then, Xu et.al.,~\cite{XuetalMaxEnt2012} find the human system in laboratory fixed-paired two-person constant-sum 2 $\times$ 2 games obey the MaxEnt. To the best of our knowledge, these are almost the total experimental works related to entropy or MaxEnt in social research field till now. In the first experimental investigation in MaxEnt, Xu et.al.~\cite{XuetalMaxEnt2012} focus on the \textbf{static} observable --- distribution and the entropy.
A direct one-step-forward question is, in the experimental social interaction systems, whether the \textbf{dynamic} observable fits MaxEnt or not?

Answering this question is the main aim of this report.
The paper is organized as follow: section two describes the relative notions; section three  introduces the experiments and reports the experimental social transitions; section four provides the MaxEnt prediction relating to the social transitions of the investigated experiments; section five reports the results; Discussion and summary are at last.

\section{Relative Notions}

\subsection{Two person constant sum $2 \times 2$ game}

Two-person zero-sum games describe situations in which two individuals are absolutely opposite to each other, where one's gain is always the other's loss~\cite{myerson1997game}. Constant sum game is strategically equivalent to zero sum games in mathematical view.

 \begin{figure}
\centering
\includegraphics[angle=0,width=3.5cm]{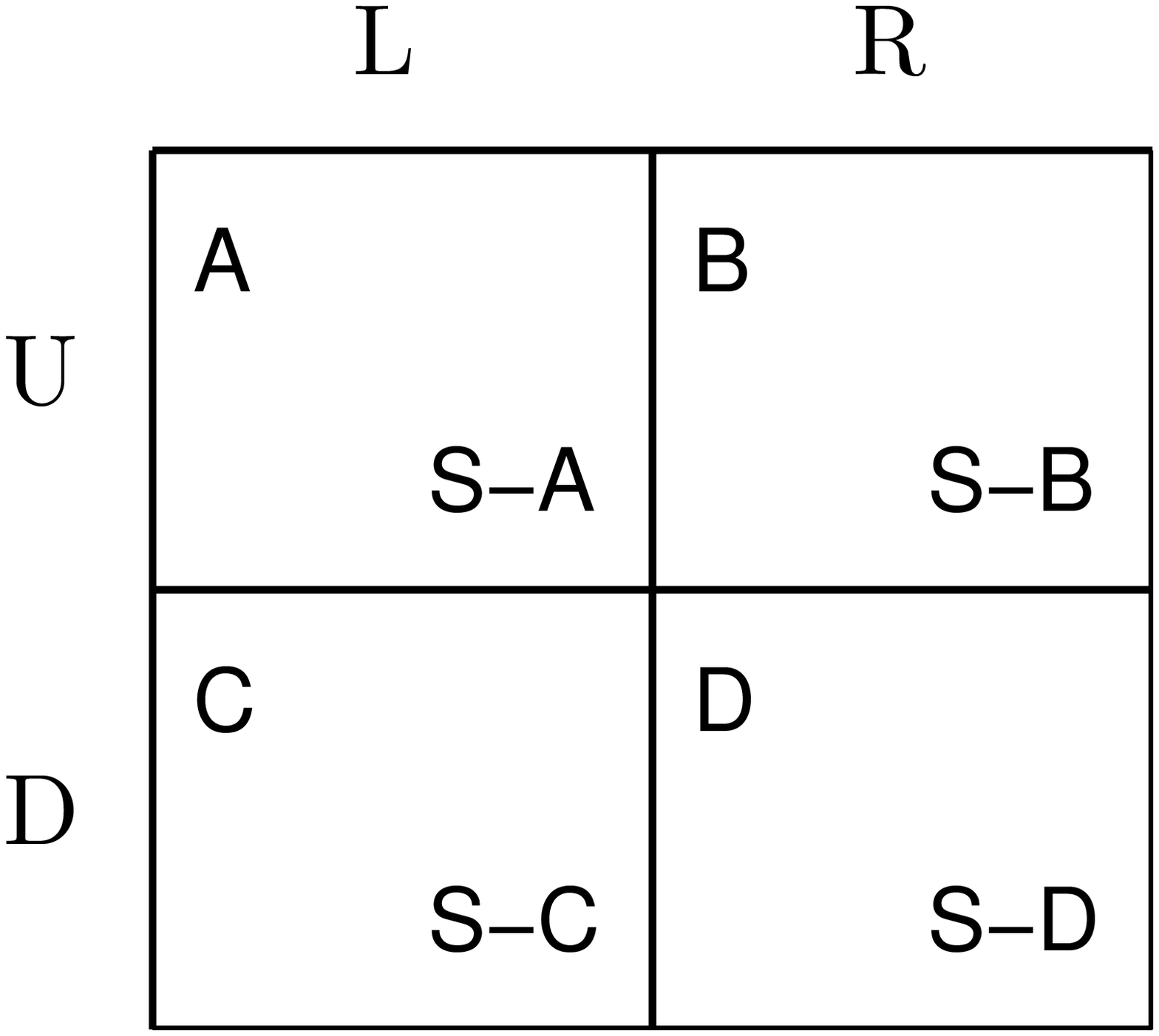}
\caption{Payoff Matrix of a Constant Sum Game.  A, B, C  and  D are the payoffs for row player under four combinations of two player's strategies respectively and S-A, S-B, S-C and S-D are for column player respectively. S denotes the sum of the payoffs to the two players.
\label{fig:payoffmatrix}}
\end{figure}

In a two-person constant-sum game, each player has two strategies. For a row player, the strategy set is $(U, D)$ and for a column player, the strategy set is $(L, R)$. The sum of the two players' payoffs is the same for any outcome. Let S denote the sum of the payoffs of the two players. Any constant sum 2 $\times$ 2 game can be written in the  form of Fig.~\ref{fig:payoffmatrix}. A, B, C  and  D is the payoff for row player under four combinations of two players' strategies respectively and S-A, S-B, S-C and S-D are for column player respectively. If $(A>C) \cap (B<D) \cap (A>B) \cap (C<D) $ or $(A<C) \cap (B>D) \cap (A<B) \cap (C>D)$ as in~\cite{selten2008}, there exists an unique mixed strategy Nash equilibrium (MSNE).

\subsection{Social State and Observation}

The social state $x_{ij}$~\cite{Sandholm2009Encyclopedia} can be taken as the combination of two players' strategies, herein $i$ indicates the column player's strategy and $j$ indicates the row player's strategy. Let $p$ be the probability of strategy R for column player  and $q$ be the probability of strategy D for row player, the social state can be described by $x_{ij}$=$(p, q)$. During a game, each player chooses a pure strategy from his own strategy set in a round $t$, the combination of these two strategies can be taken as a \textbf{social state}  $x_{t}$ in that round, $x_t$ = $(p_t, q_t)$. Obviously, there are altogether four possible social states in one round, i.e., (0,0), (0,1), (1,0)  and (1,1) indicating LU, LD, RU  and RD respectively, and we simplify them as $x_{00}$, $x_{01}$, $x_{10}$, and $x_{11}$. In Fig.~\ref{fig:Rothquivershiyi}(b) and (f), the gray dots present the social states.

If the game is repeated, \textbf{observation} denoted as $\Omega_{ij}$ at each state $x_{ij}$ can be accumulated and the results of these games are shown in the last 4 columns of Tab~\ref{tab:RothPayoffEntropy}.

\subsection{Social transition}\label{socTr}

In this paper, we investigate the social transitions within the strategy states in the strategy space.
In a repeated game, for a given round $t$, the social state is $x_{t}=(p_t, q_t)$; similarly, the social state in the previous round can be denoted as $x_{t-1}:=(p_{t-1}, q_{t-1})$ and  in the next round can be denoted as $x_{t+1}:= (p_{t+1}, q_{t+1})$. For each given round $t$, there exists the next round and previous round except the first round and last round in a experimental session. So, there exists a social \textbf{forward transition} vector (denoted as $T_+$)  indicating the transition from $x_t$ to $x_{t+1}$, and a social \textbf{backward transition} vector (denoted as $T_-$) indicating the transition from $x_{t-1}$ to $x_t$.

In a two-person 2$\times$2 game, there are four social states,
so there are all 32 transitions (shown in the first column in Table~\ref{tab:Actual Frequencies}), including the 4 backward (forward) transitions for each of the 4 states.
These 32 transitions should be the samples for MaxEnt testing.
%
\subsection{Distribution of Transitions of a Given State}\label{disTr}

During a game, for a given social state, there exists 4 forward transitions and 4 backward transitions, respectively. This means that there should exist a distribution of transitions.

\begin{figure}
\centering
\includegraphics[angle=0,width=3cm]{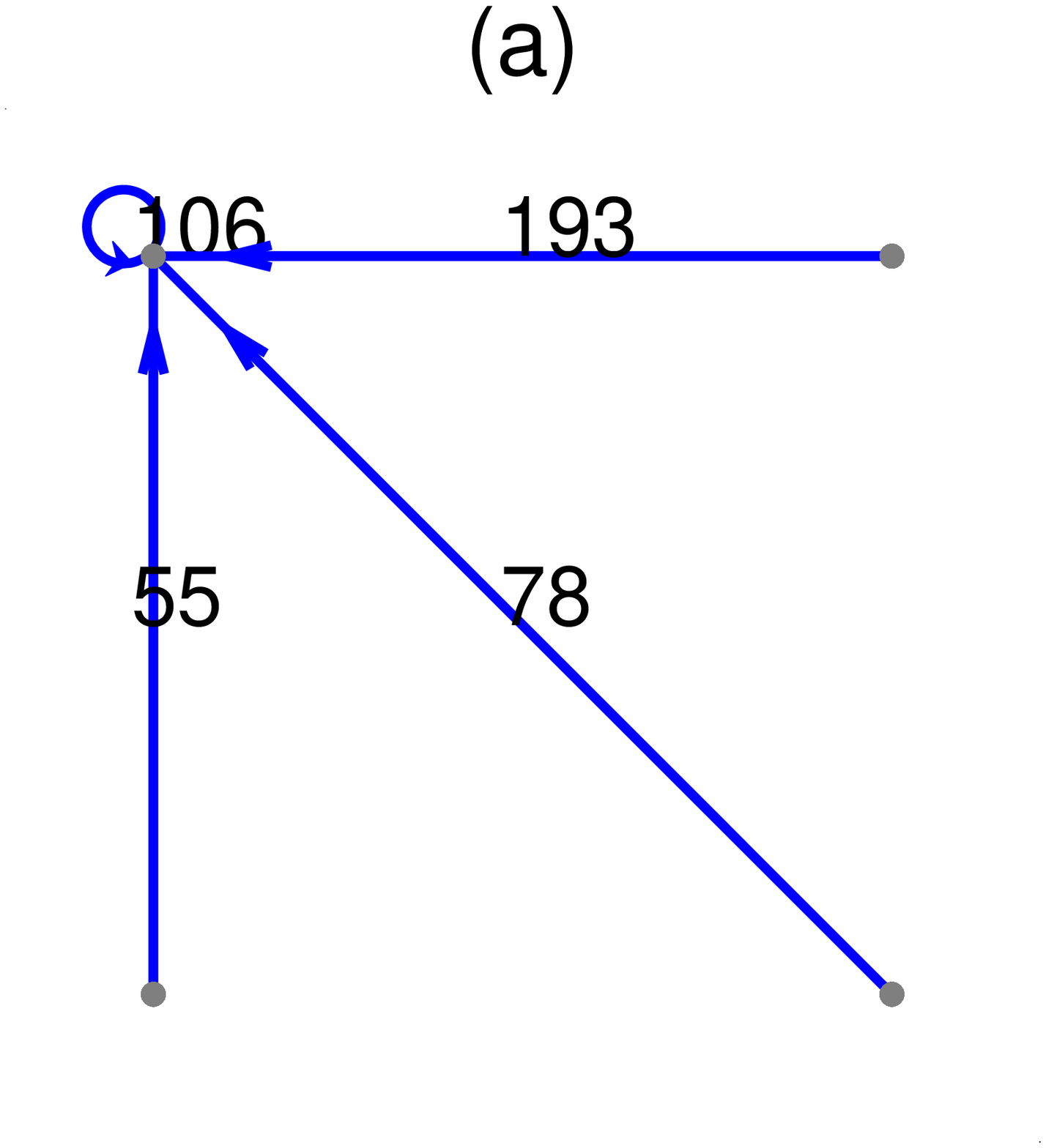}~~
\includegraphics[angle=0,width=3cm]{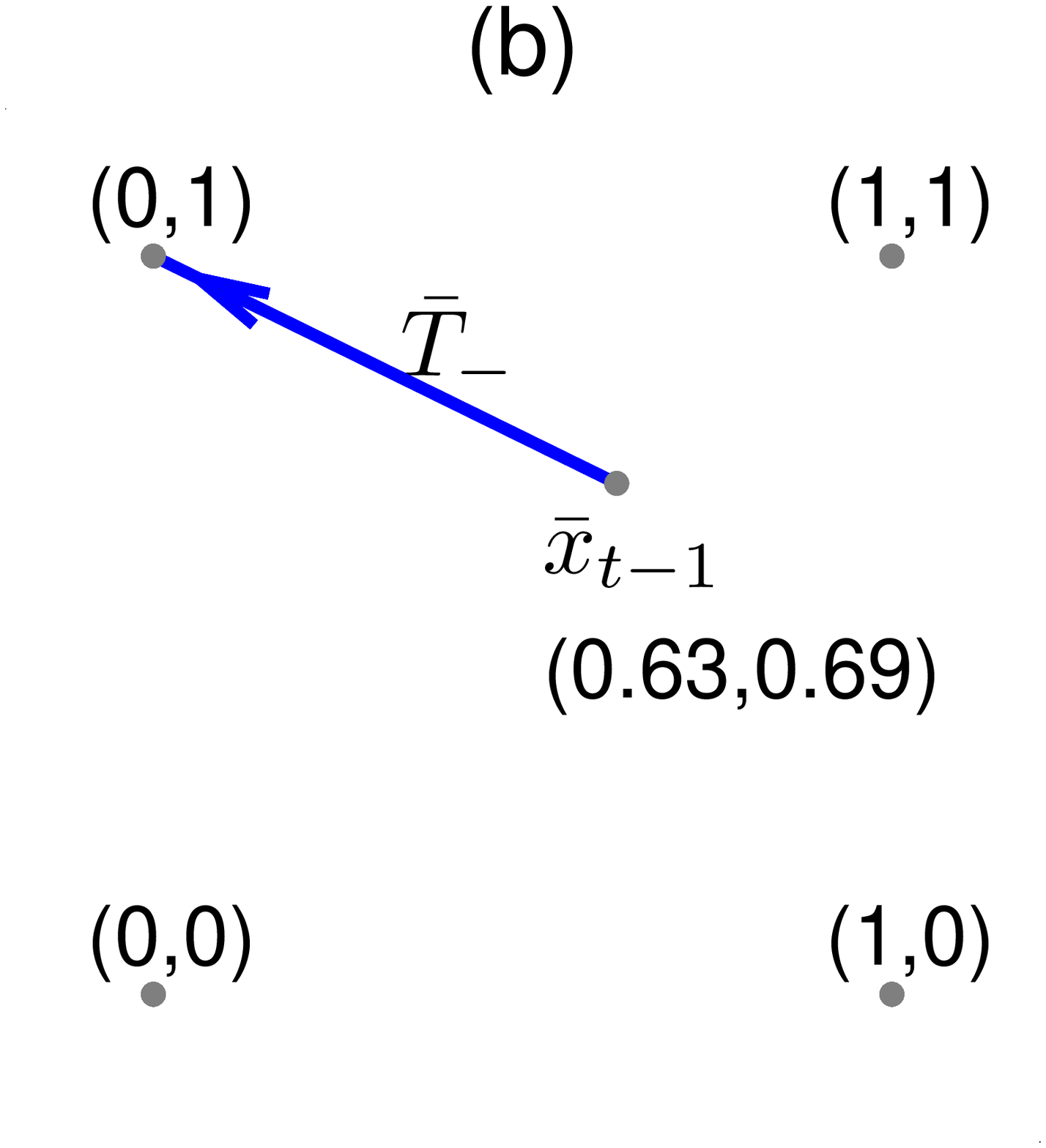}~~
\includegraphics[angle=0,width=3cm]{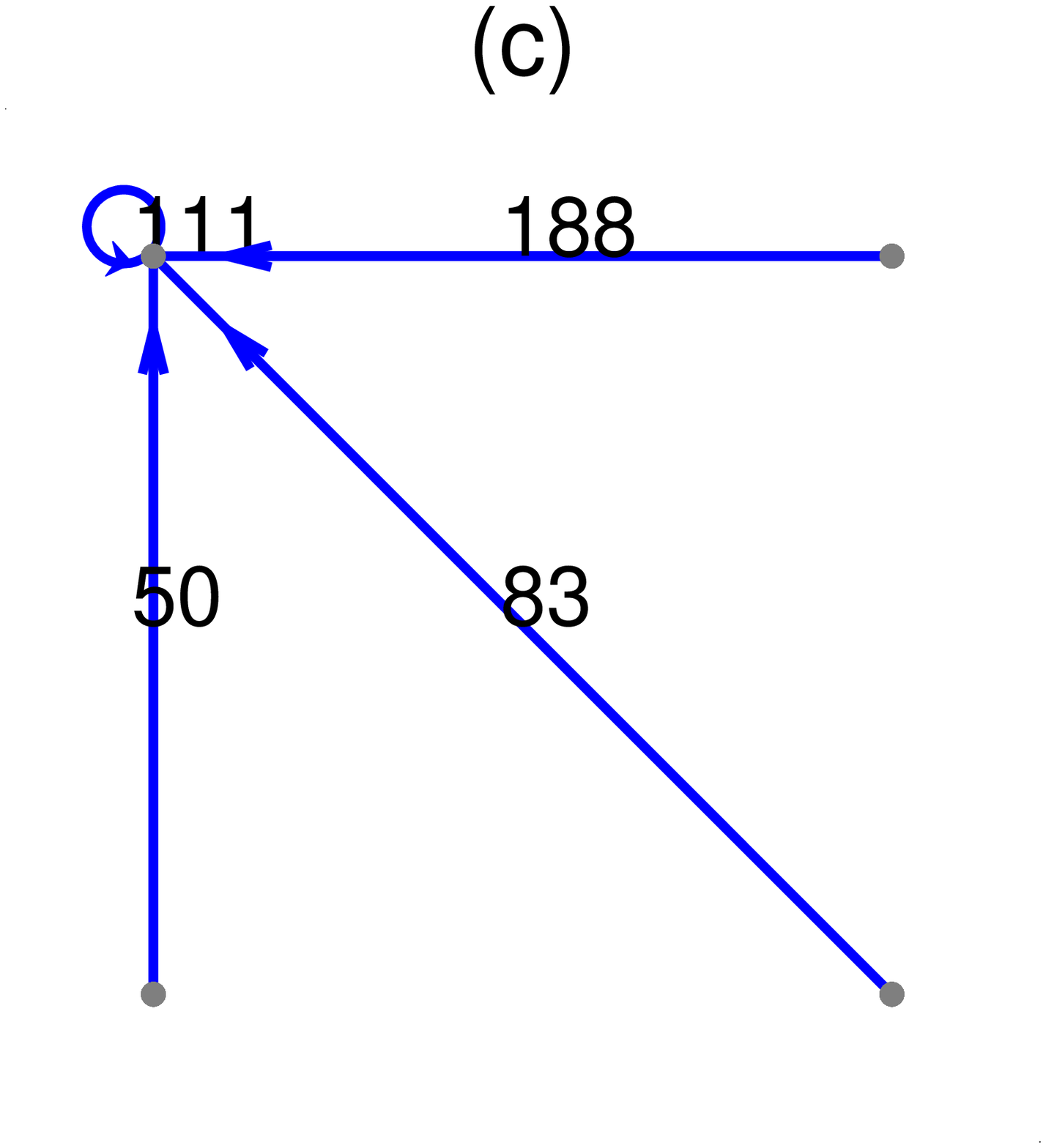}~~
\includegraphics[angle=0,width=3cm]{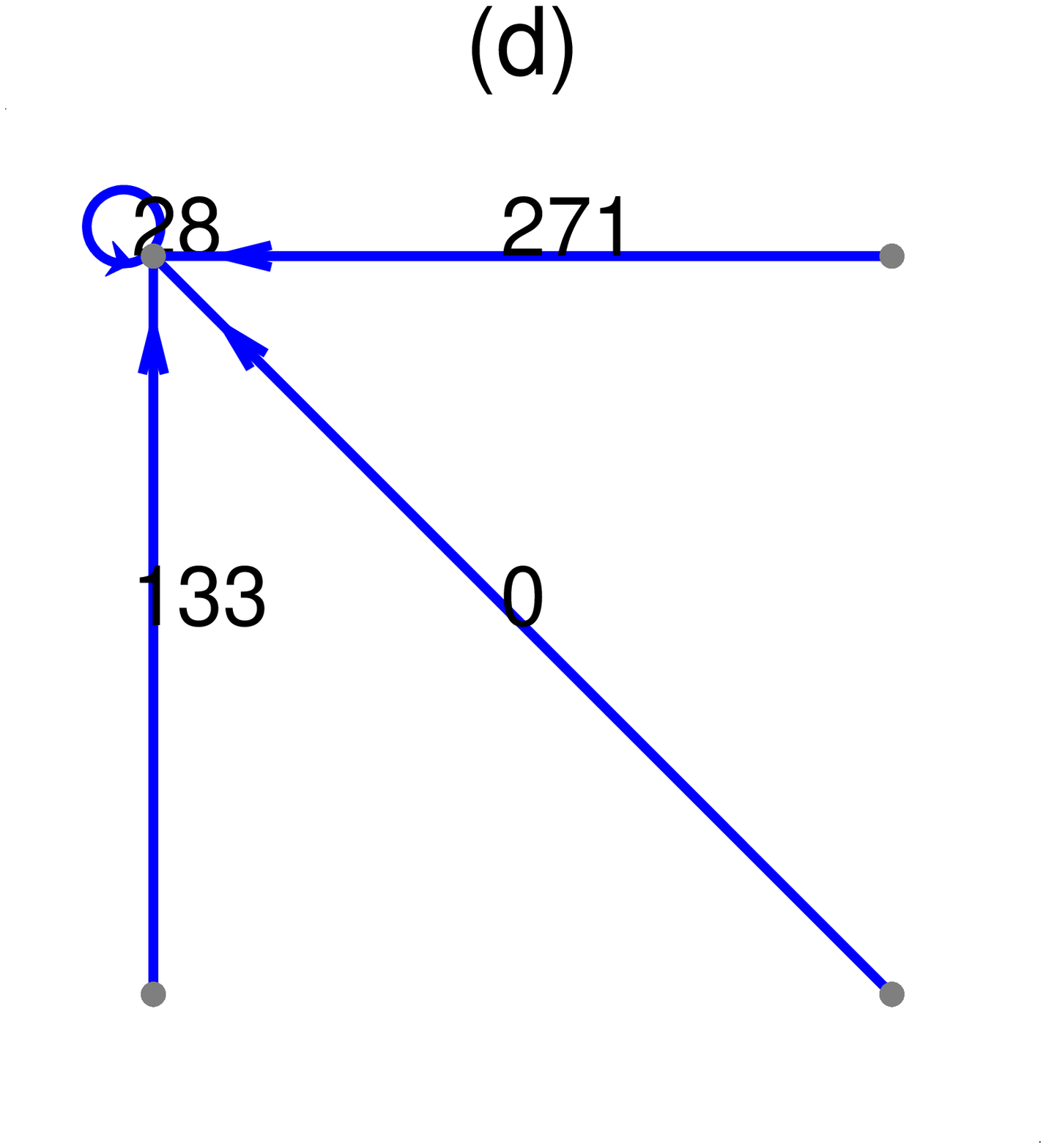}\\
\includegraphics[angle=0,width=3cm]{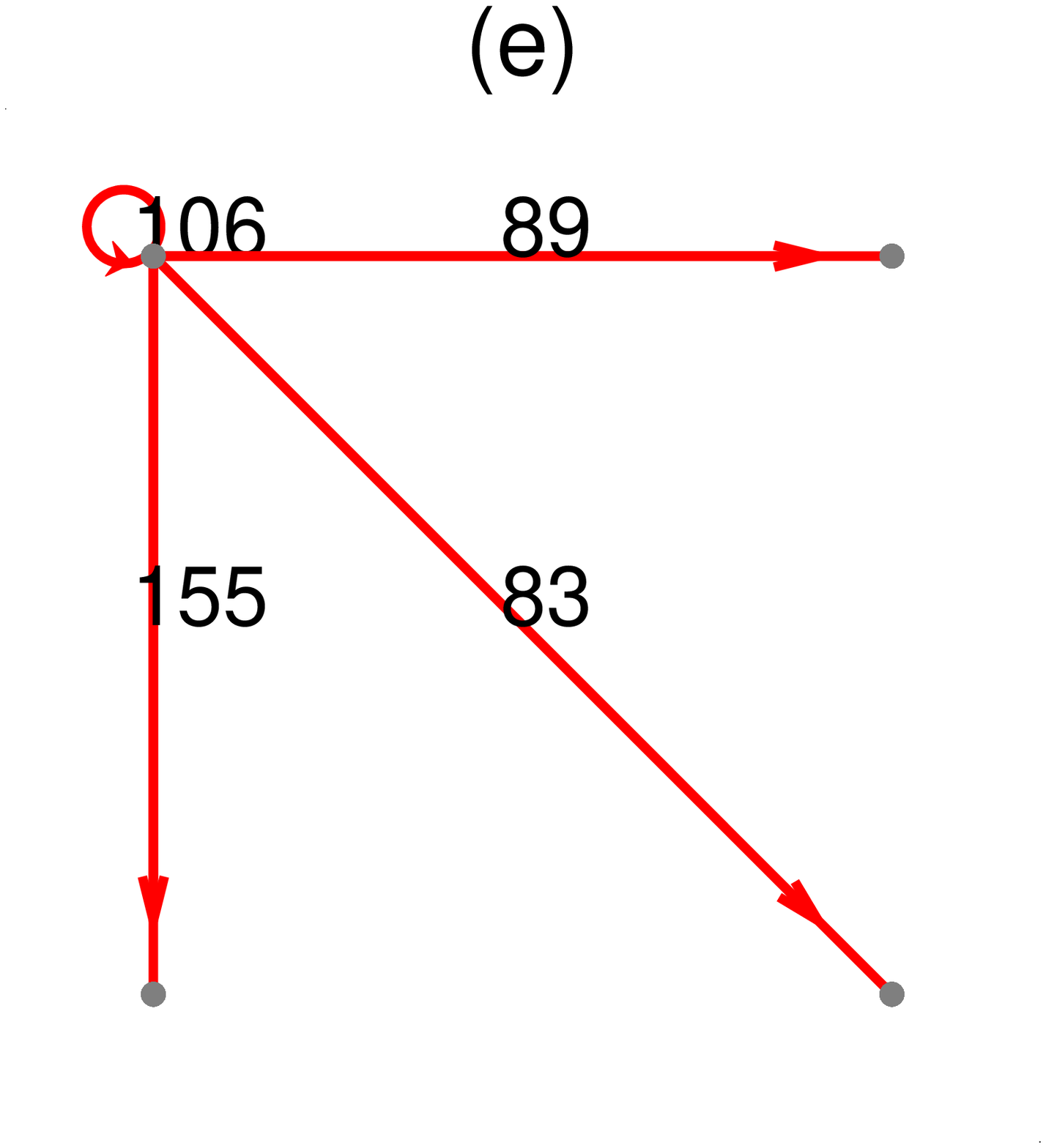}~~
\includegraphics[angle=0,width=3cm]{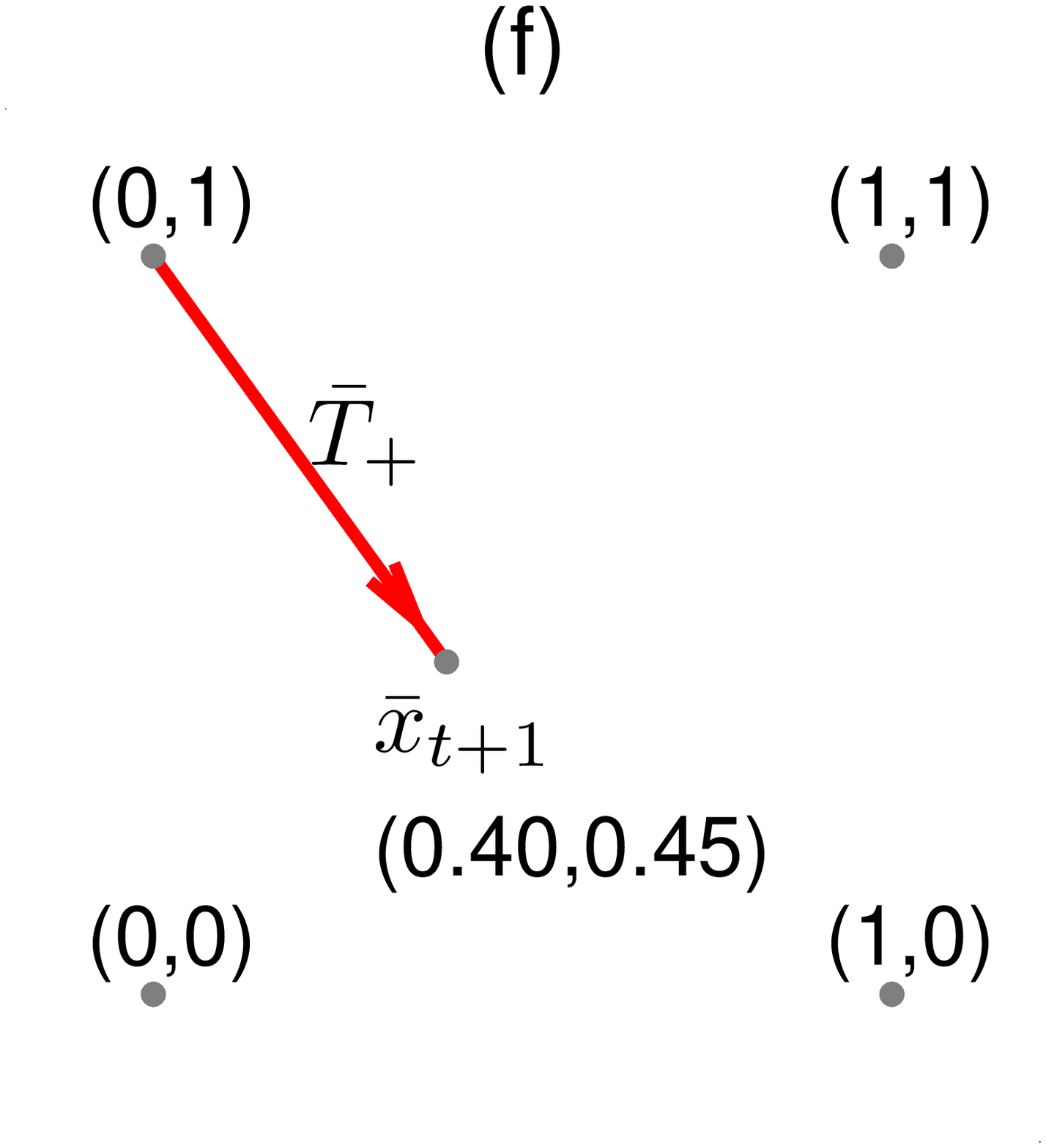}~~
\includegraphics[angle=0,width=3cm]{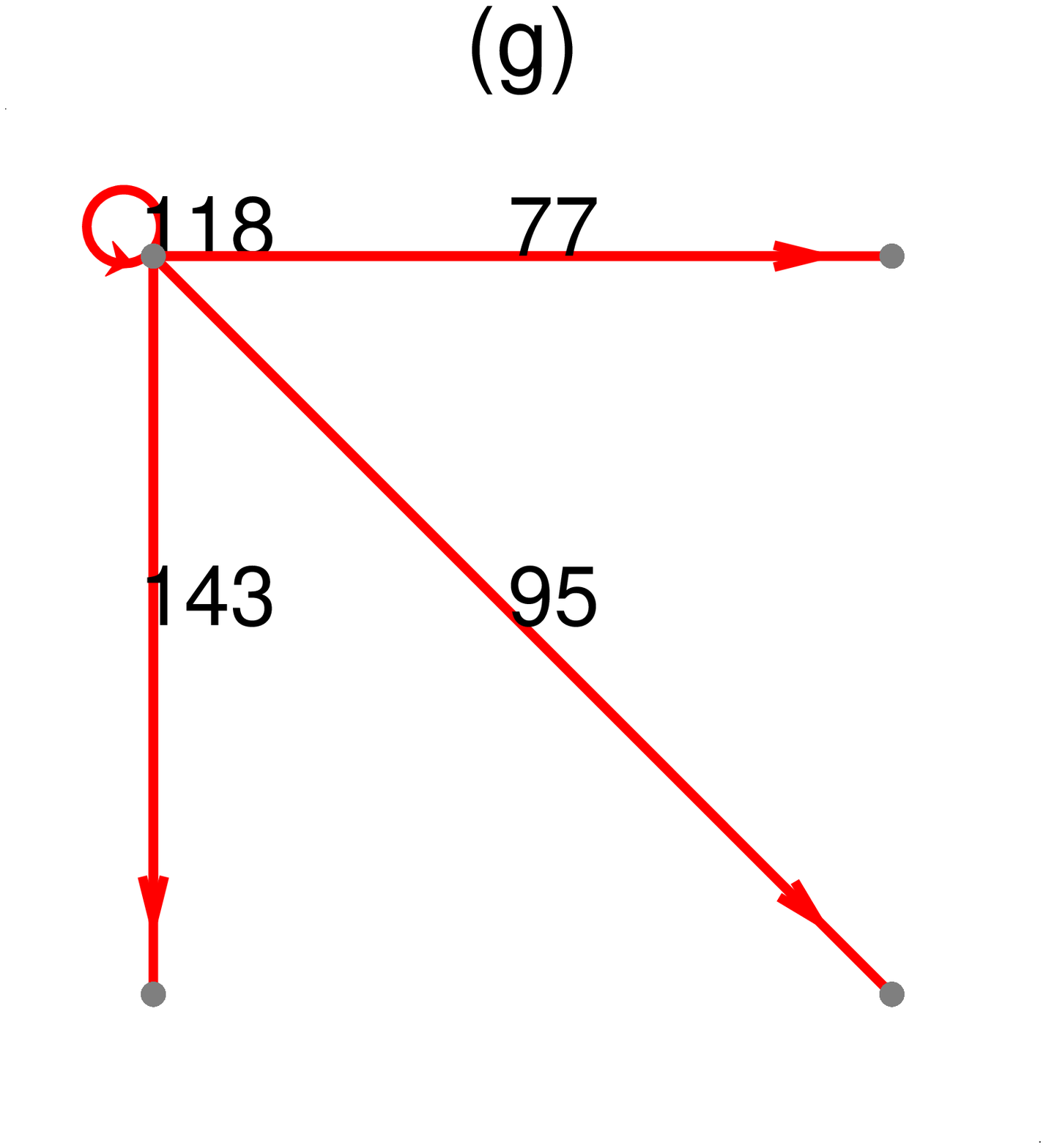}~~
\includegraphics[angle=0,width=3cm]{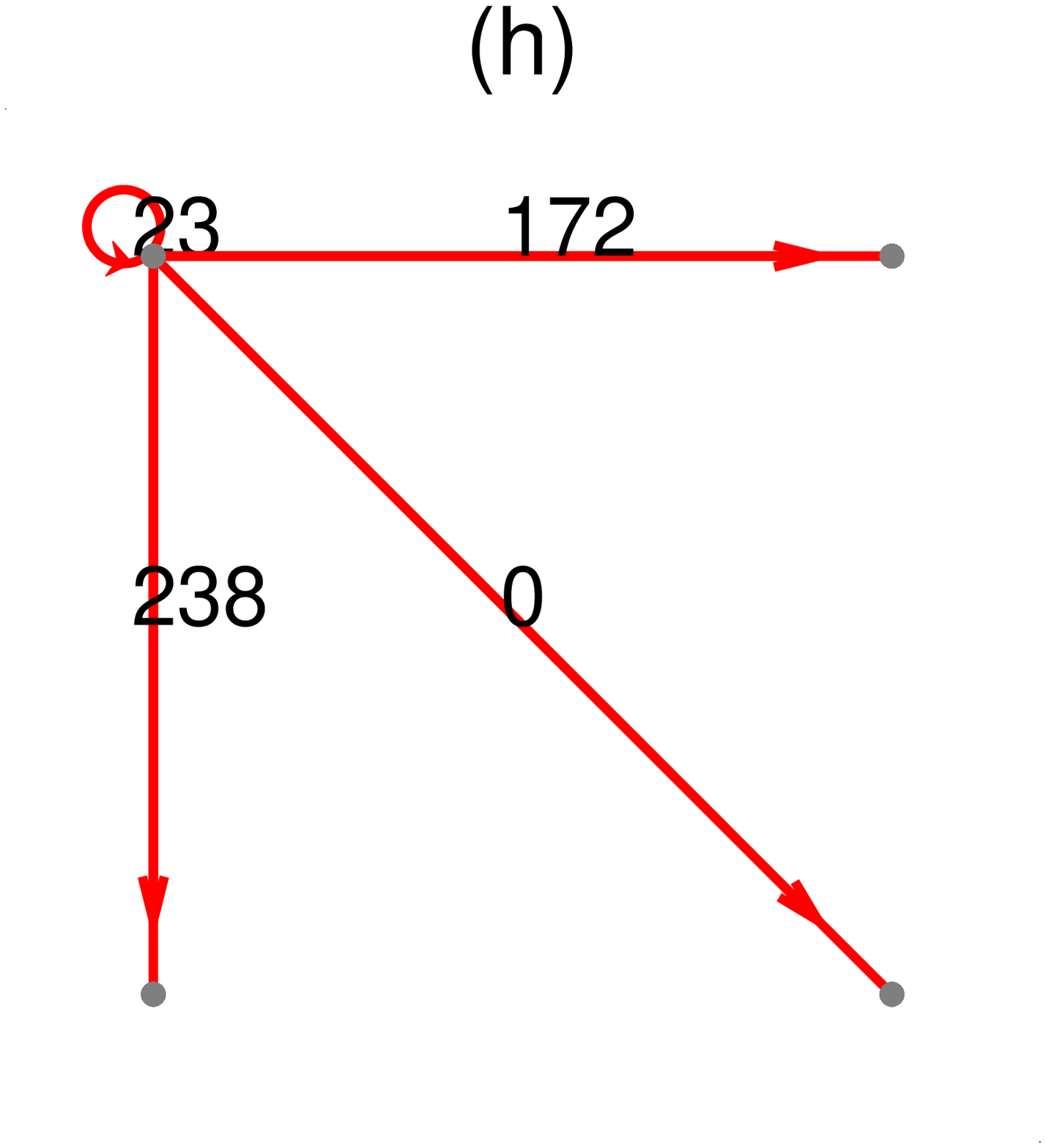}
\caption{Transition distribution and aggregated transition. (a) four actual backward transitions $T_{01{\leftarrow}00}$, $T_{01{\leftarrow}01}$,$T_{01{\leftarrow}10}$, $T_{01{\leftarrow}11}$ for the given state $x_{01}$ from $x_{00}$, $x_{01}$, $x_{10}$, $x_{11}$; the blue arrows indicate the directions of transitions; the numerics indicate the frequencies of transitions, respectively; (b) the aggregated backward transition $\bar{T}_-$ and related mean starting point ${\bar{x}_{t-1}}$ = ${(\bar{p}_{t-1}, \bar{q}_{t-1})}$ = $(0.63,0.69)$; (c) four expected backward transitions for   $x_{01}$ from four starting points $x_{0 0}$, $x_{01}$, $x_{10}$, $x_{11}$ with MaxEnt, the blue arrows indicate the directions of transitions and the numerics indicate the related frequencies;
 (d) an example which is not in agreement with MaxEnt but satisfies the aggregated backward transition constraint in (b);(e) four actual forward transitions $T_{01{\rightarrow}00}$, $T_{01{\rightarrow}01}$, $T_{01{\rightarrow}10}$, $T_{01{\rightarrow}11}$ from $x_{01}$,
  the red arrows indicate the directions of transitions, and the numerics indicate the related actual frequencies of transitions;
  (f) the aggregated forward transition $\bar{T}_+$ and relative mean terminal point ${\bar{x}_{t+1}}$ = $({\bar{p}_{t+1}}, {\bar{q}_{t+1}})$ = $(0.40,0.45)$; (g) four MaxEnt expected forward transitions; 
  (h) an example which is not in agreement with MaxEnt but satisfies the actual aggregated forward transition constraint. Data in (a) and (e) comes from the experiment --- game 1.}
\label{fig:Rothquivershiyi}
\end{figure}

For example, Fig.~\ref{fig:Rothquivershiyi} (a) is demonstrating the distribution of the transitions of the given state $x_{01}$, in which the four backward transitions $T_{01{\leftarrow}00}$, $T_{01{\leftarrow}01}$,$T_{01{\leftarrow}10}$ and $T_{01{\leftarrow}11}$  come from the four state $x_{00}$, $x_{01}$, $x_{10}$ and $x_{11}$, respectively;  the blue arrows indicate the directions of transitions and the numerics indicate the related actual frequencies. The distribution of backward transitions $T_{01{\leftarrow}00}$, $T_{01{\leftarrow}01}$,$T_{01{\leftarrow}10}$, $T_{01{\leftarrow}11}$ are 55, 106, 78, and 193, respectively. Similarly, Fig.~\ref{fig:Rothquivershiyi} (e) illustrates the distribution of  the forward transitions.

\subsection{Aggregated Transition of a State}\label{aggTr}

The existence of distribution of transitions of a given state implicates that there are many backward starting points and forward terminal points. So, for a given state $x_{i_0j_0}$, we can get a so-called the mean starting point ${\bar{x}_{t-1}}$=${(\bar{p}_{t-1}, \bar{q}_{t-1})}$ and a \textbf{aggregated backward transition} $\bar{T}_-$ (it is natural that $\bar{T}_-$=${x_{i_0j_0} - \bar{x}_{t-1}}$), and also the mean terminal point ${\bar{x}_{t+1}}$ = $({\bar{p}_{t+1}}, {\bar{q}_{t+1}})$ and a \textbf{aggregated forward transition} $\bar{T}_+$ (it is natural that $\bar{T}_+$=${\bar{x}_{t+1}} - x_{i_0j_0}$). The aggregated forward transition $T_+$ is the same as the experimental dynamics observable in literatures (called as change in a given state in ref.~\cite{HuyckSamuelson2001} and the mean jump-out vector of a given state in ref~\cite{XuWang2011ICCS}).

For example, supposing the given state is (0,1), Fig.~\ref{fig:Rothquivershiyi} (b) illustrates the aggregated backward transition $\bar{T}_-$, as the average of the four vectors in Fig.~\ref{fig:Rothquivershiyi} (a), is (-0.63,0.31); Meanwhile the mean starting point ${\bar{x}_{t-1}}$=$(0.63,0.69)$; In Fig.~\ref{fig:Rothquivershiyi} (f),  $\bar{T}_+$, as the average of the four vectors in Fig.~\ref{fig:Rothquivershiyi} (e), is (0.40,-0.55) and then  ${\bar{x}_{t+1}}$=$(0.40,0.45)$.

\section{Data Set and Experimental Transitions}
\subsection{Experiments and Data set}
Experimental economics methods are well suited to evaluate theories~\cite{Falk2009}. In this paper, we use the same data set as ref~\cite{XuetalMaxEnt2012} to test the MaxEnt in social strategy transitions. The two-person constant sum 2$\times$2 game includes 11 different parameters (Table~\ref{tab:RothPayoffEntropy}).  
From game 1 to game 10, each game consists of 9 pairs of subjects, each pair play for 500 rounds while for game 11, the game consists of 12 pairs of subjects, each pair play for 300 rounds. These yield 4500 observed social states in each of game 1 to game 10 and 3600 observed social states in game 11 (for more detail, see ref.~\cite{RothErev2007,XuetalMaxEnt2012}).

\begin{table}[htbp2]
\centering
\begin{threeparttable}
\small
\caption{\label{tab:RothPayoffEntropy} Parameters and observations of the 11 game$^\S$}
\begin{tabular}{|c|ccccc|cc|rrrr|}
  \hline
   \hline
   game	&A	&B	&C	&D	&S	&Group	&Rounds	&$\Omega_{00}$	&$\Omega_{01}$	&$\Omega_{10}$		 &$\Omega_{11}$	\\
   \hline
~g1	&77	&35	&~8	&48	&100	&~9	&500	&994 &433	&1659	 &1405	\\
~g2	&73	&74	&87	&20	&100	&~9	&500	&1373&250	&2401	 &467	\\
~g3	&63	&~8	&~1	&17	&100	&~9	&500	&664 &333	&1955	 &1539	\\
~g4	&55	&75	&73	&60	&100	&~9	&500	&643 &1611	&588	 &1649	\\
~g5	&~5	&64	&93	&40	&100	&~9	&500	&548 &891	&1153	 &1899	\\
~g6	&46	&54	&61	&23	&100	&~9	&500	&1135 &706	&1729	 &921	\\
~g7	&89	&53	&82	&92	&100	&~9	&500	&502 &1840	&825	 &1324	\\
~g8	&88	&38	&40	&55	&100	&~9	&500	&353 &663	&1443	 &2032	\\
~g9	&40	&76	&91	&23	&100	&~9	&500	&1157	&860	 &1366	&1108	\\
g10	&69	&~5	&13	&33	&100	&~9	&500	&443	&465	&995	 &2588	\\
g11	&~5	&~0	&~0	&~5	&~~5	&12	&300	&837	&913	&907	 &931	\\
  \hline
    \hline
\end{tabular}
\begin{tablenotes}
  \item[$\S$] g1  to  g11 indicate game 1 to game 11, respectively. The symbols A, B, C, D and S refer to Fig.~\ref{fig:payoffmatrix}. Group is the number of the pairs of human subjects playing the games. Rounds is the game repeated times in each pair.
  \end{tablenotes}
\end{threeparttable}
\end{table}

\subsection{Experimental Distributions of Transitions}

According to the data set for each of the 11 games, using the definition in Sec. \ref{socTr} and \ref{disTr},
we can calculate the actual experimental distributions of backward transitions and  forward transitions. The results are summarized in Table~\ref{tab:Actual Frequencies}, which
should serve as the targets for testing MaxEnt.

\subsection{Experimental Aggregated Transition for Each State}
%

Numerically, $\bar{x}_{t\pm1}$ can be presented by two components ($\bar{p}_{t\pm1},\bar{q}_{t\pm1}$). Using the definition in Sec.~\ref{aggTr},
results of all of the components from experiments are shown  in Table~\ref{tab: mean start points and terminal points }. The vectors, $\bar{x}_{t\pm1}$, are shown in the sub-figures in Fig.~\ref{fig:roth}.

For calculating the theoretical backward (forward) distributions of the transitions from MaxEnt, $\bar{x}_{t-1}$ ($\bar{x}_{t+1}$) should constraint the testing of MaxEnt.

%

\begin{table}[htbp2]
\centering
\begin{threeparttable}
\small
\caption{\label{tab:Actual Frequencies} Actual frequencies of the transitions of 11 games}
\begin{tabular}{|c|ccccccccccc|}
  \hline
   \hline
$T_-$ &~~g1	&~~g2	&~~g3	&~~g4	&~~g5	&~~g6	&~~g7	&~~g8	&~~g9	&~g10	&~g11	\\
  \hline
$T_{00{\leftarrow}00}$	&~464	&~764	&~184	&~314	&~124	&~529	&~143	&~111	&~606	 &~116	 &~196	\\
$T_{00{\leftarrow}01}$	&~155	&~~52	&~~73	&~~67	&~~68	&~~99	&~169	&~~95	&~~67	 &~139	 &~241	\\
$T_{00{\leftarrow}10}$	&~274	&~504	&~327	&~193	&~207	&~382	&~~89	&~~70	&~362	 &~123	 &~182	\\
$T_{00{\leftarrow}11}$	&~102	&~~53	&~~79	&~~66	&~149	&~124	&~~98	&~~75	&~120	 &~~63	 &~218	\\
$T_{01{\leftarrow}00}$	&~~55	&~~86	&~~35	&~~239	&~213	&~226	&~104	&~~11	&~263	 &~~43	 &~149	\\
$T_{01{\leftarrow}01}$	&~106	&~~48	&~~89	&1054	&~217	&~311	&1191	&~264	&~365	 &~121	 &~216	\\
$T_{01{\leftarrow}10}$	&~~78	&~~69	&~100	&~~70	&~191	&~~86	&~~66	&~~62	&~~85	 &~~55	 &~231	\\
$T_{01{\leftarrow}11}$	&~193	&~~45	&~111	&~245	&~268	&~~85	&~482	&~327	&~145	 &~247	 &~319	\\
$T_{10{\leftarrow}00}$	&~401	&~446	&~383	&~~45	&~~51	&~235	&~145	&~169	&~144	 &~232	 &~281	\\
$T_{10{\leftarrow}01}$	&~~83	&~~75	&~~99	&~~65	&~143	&~~86	&~~99	&~~82	&~~99	 &~~91	 &~263	\\
$T_{10{\leftarrow}10}$	&1021	&1722	&1046	&~258	&~483	&1029	&~478	&~858	&~691	 &~370	 &~191	\\
$T_{10{\leftarrow}11}$	&~152	&~160	&~424	&~223	&~476	&~380	&~103	&~333	&~434	 &~302	 &~173	\\
$T_{11{\leftarrow}00}$	&~~74	&~~77	&~~62	&~~45	&~160	&~145	&~110	&~~62	&~144	 &~~52	 &~211	\\
$T_{11{\leftarrow}01}$	&~~89	&~~75	&~~72	&~425	&~463	&~210	&~381	&~222	&~329	 &~114	 &~193	\\
$T_{11{\leftarrow}10}$	&~286	&~106	&~482	&~~67	&~272	&~232	&~192	&~453	&~228	 &~447	 &~303	\\
$T_{11{\leftarrow}11}$	&~958	&~209	&~925	&1115	&1006	&~332	&~641	&1297	&~409	&1976	 &~221	\\
  \hline
$T_+$&~~g1	&~~g2	&~~g3	&~~g4	&~~g5	&~~g6	&~~g7	&~~g8	&~~g9	&~g10	&~g11	\\
  \hline
$T_{00{\rightarrow}00}$	&~464	&~764	&~184	&~314	&~124	&~529	&~143	&~111	&~606	 &~116	 &~196	\\
$T_{00{\rightarrow}01}$	&~~55	&~~86	&~~35	&~239	&~213	&~226	&~104	&~~11	&~263	 &~~43	 &~149	\\
$T_{00{\rightarrow}10}$	&~401	&~446	&~383	&~~45	&~~51	&~235	&~145	&~169	&~144	 &~232	 &~281	\\
$T_{00{\rightarrow}11}$	&~~74	&~~77	&~~62	&~~45	&~160	&~145	&~110	&~~62	&~144	 &~~52	 &~211	\\
$T_{01{\rightarrow}00}$	&~155	&~~52	&~~73	&~~67	&~~68	&~~99	&~169	&~~95	&~~67	 &~139	 &~241	\\
$T_{01{\rightarrow}01}$	&~106	&~~48	&~~89	&1054	&~217	&~311	&1191	&~264	&~365	 &~121	 &~216	\\
$T_{01{\rightarrow}10}$	&~~83	&~~75	&~~99	&~~65	&~143	&~~86	&~~99	&~~82	&~~99	 &~~91	 &~263	\\
$T_{01{\rightarrow}11}$	&~~89	&~~75	&~~72	&~425	&~463	&~210	&~381	&~222	&~329	 &~114	 &~193	\\
$T_{10{\rightarrow}00}$	&~274	&~504	&~327	&~193	&~207	&~382	&~~89	&~~70	&~362	 &~123	 &~182	\\
$T_{10{\rightarrow}01}$	&~~78	&~~69	&~100	&~~70	&~191	&~~86	&~~66	&~~62	&~~85	 &~~55	 &~231	\\
$T_{10{\rightarrow}10}$	&1021	&1722	&1046	&~258	&~483	&1029	&~478	&~858	&~691	 &~370	 &~191	\\
$T_{10{\rightarrow}11}$	&~286	&~106	&~482	&~~67	&~272	&~232	&~192	&~453	&~228	 &~447	 &~303	\\
$T_{11{\rightarrow}00}$	&~102	&~~53	&~~79	&~~66	&~149	&~124	&~~98	&~~75	&~120	 &~~63	 &~218	\\
$T_{11{\rightarrow}01}$	&~193	&~~45	&~111	&~245	&~268	&~~85	&~482	&~327	&~145	 &~247	 &~319	\\
$T_{11{\rightarrow}10}$	&~152	&~160	&~424	&~223	&~476	&~380	&~103	&~333	&~434	 &~302	 &~173	\\
$T_{11{\rightarrow}11}$	&~958	&~209	&~925	&1115	&1006	&~332	&~641	&1297	&~409	 &1976	 &~221	\\
  \hline
    \hline
\end{tabular}
\begin{tablenotes}
  \item[$\S$] g1 indicates game 1, the rest analogize.
  \end{tablenotes}
\end{threeparttable}
\end{table}

\section{Theoretical Distributions of Transitions from MaxEnt}

In this paper, in order to have a deeper insight in the dynamic social observable, we use the aggregated social transitions ($T_\pm$)
as the constraints for MaxEnt testing.
It is clear that for a given state, without MaxEnt, the distribution of backward (forward) transitions can be \textbf{arbitrary} even given the constraints $T_\pm$ (two examples are given in discussion).



\begin{table}[htbp2]
\centering
\begin{threeparttable}
\small
\caption{\label{tab: mean start points and terminal points } Mean starting point $\bar{x}_{t-1}$ and terminal point $\bar{x}_{t+1}$ }
\begin{tabular}{|cc|ccccccccccc|}
  \hline
   \hline
state	&${\bar{x}_{t-1}}$	&g1	&g2	&g3	&g4	&g5	&g6	&g7	&g8	&g9	&g10	&g11	\\
  \hline
$x_{00}$	&${\bar{p}_{t-1}}$	&0.38	&0.41	&0.61	&0.40	&0.65	&0.45	&0.37	&0.41	 &0.42	 &0.42	 &0.48	 \\
 	&${\bar{q}_{t-1}}$	&0.26	&0.08	&0.23	&0.21	&0.40	&0.20	&0.54	&0.48	 &0.16	 &0.46	 &0.55	\\
$x_{01}$	&${\bar{p}_{t-1}}$	&0.63	&0.46	&0.63	&0.20	&0.52	&0.24	&0.30	&0.59	 &0.27	 &0.65	 &0.60	 \\
 	&${\bar{q}_{t-1}}$	&0.69	&0.38	&0.6	&0.81	&0.55	&0.56	&0.91	&0.89	 &0.59	 &0.79	 &0.58	\\
$x_{10}$	&${\bar{p}_{t-1}}$	&0.71	&0.78	&0.75	&0.81	&0.83	&0.81	&0.70	&0.83	 &0.82	 &0.68	 &0.40	 \\
 	&${\bar{q}_{t-1}}$	&0.14	&0.10	&0.27	&0.49	&0.54	&0.27	&0.24	&0.29	 &0.39	 &0.39	 &0.48	\\
$x_{11}$	&${\bar{p}_{t-1}}$	&0.88	&0.67	&0.91	&0.72	&0.67	&0.61	&0.63	&0.86	 &0.57	 &0.94	 &0.56	 \\
 	&${\bar{q}_{t-1}}$	&0.74	&0.61	&0.65	&0.93	&0.77	&0.59	&0.77	&0.75	 &0.66	 &0.81	 &0.45	\\
  \hline
state	&${\bar{x}_{t+1}}$	&g1	&g2	&g3	&g4	&g5	&g6	&g7	&g8	&g9	&g10	&g11	\\
  \hline
$x_{00}$	&${\bar{p}_{t+1}}$	&0.48	&0.38	&0.67	&0.14	&0.39	&0.33	&0.51	&0.65	 &0.25	 &0.64	 &0.59	 \\
 	&${\bar{q}_{t+1}}$	&0.13	&0.12	&0.15	&0.44	&0.68	&0.33	&0.43	&0.21	 &0.35	 &0.21	 &0.43	\\
$x_{01}$	&${\bar{p}_{t+1}}$	&0.40	&0.60	&0.51	&0.30	&0.68	&0.42	&0.26	&0.46	 &0.50	 &0.44	 &0.50	 \\
 	&${\bar{q}_{t+1}}$	&0.45	&0.49	&0.48	&0.92	&0.76	&0.74	&0.85	&0.73	 &0.81	 &0.51	 &0.45	\\
$x_{10}$	&${\bar{p}_{t+1}}$	&0.79	&0.76	&0.78	&0.55	&0.65	&0.73	&0.81	&0.91	 &0.67	 &0.82	 &0.54	 \\
 	&${\bar{q}_{t+1}}$	&0.22	&0.07	&0.30	&0.23	&0.40	&0.18	&0.31	&0.36	 &0.23	 &0.5	 &0.59	\\
$x_{11}$	&${\bar{p}_{t+1}}$	&0.79	&0.79	&0.88	&0.81	&0.78	&0.77	&0.56	&0.80	 &0.76	 &0.88	 &0.42	 \\
 	&${\bar{q}_{t+1}}$	&0.82	&0.54	&0.67	&0.82	&0.67	&0.45	&0.85	&0.80	 &0.50	 &0.86	 &0.58	\\
  \hline
  \hline
\end{tabular}
\begin{tablenotes}
  \item[$\S$] g1 indicates game 1, the rest analogize.
  \end{tablenotes}
\end{threeparttable}
\end{table}


For a given state $x_{i_0j_0}$, the $\bar{x}_{t-1}$ is assumed to be $(\bar{p}_{t-1},\bar{q}_{t-1})$
and there is no other information. According to MaxEnt suggested by Jaynes~\cite{Jaynes2003}, the probability of the backward transitions from the states to $x_{i_0j_0}$ can be expressed, respectively, as
\begin{eqnarray*}
  p(T_{i_0j_0 \leftarrow 00}|\bar{x}_{t-1})&=&(1-\bar{p}_{t-1})^{1-i}  (1-\bar{q}_{t-1})^{1-j} \\
  p(T_{i_0j_0 \leftarrow 01}|\bar{x}_{t-1})&=&\bar{q}_{t-1}^j (1-\bar{p}_{t-1})^{1-i}     \\
  p(T_{i_0j_0 \leftarrow 10}|\bar{x}_{t-1})&=&\bar{p}_{t-1}^i (1-\bar{q}_{t-1})^{1-j} \\
  p(T_{i_0j_0 \leftarrow 11}|\bar{x}_{t-1})&=&\bar{p}_{t-1}^i  \bar{q}_{t-1}^j
\end{eqnarray*}
More compactly, the probability of backward transitions from $x_{ij}$ to $x_{i_0j_0}$, can be expressed as,
 \begin{equation}\label{eq:a4}
     p(T_{i_0j_0 \leftarrow ij}|\bar{x}_{t-1}) = \bar{p}_{t-1}^i(1-\bar{p}_{t-1})^{1-i}  \bar{q}_{t-1}^j(1-\bar{q}_{t-1})^{1-j} ,
\end{equation}
in which $\{i,j\} \in \{0, 1\}$.
Similarly, for $\bar{x}_{i_{0}j_{0}}$ and its $\bar{x}_{t+1}$=($\bar{p}_{t+1},\bar{q}_{t+1}$), the probability of the forward transition from $x_{i_0j_0}$ to state $x_{ij}$ can be expressed as
 \begin{equation}\label{eq:a5}
     p(T_{i_0j_0 \rightarrow ij}|\bar{x}_{t+1}) = \bar{p}_{t+1}^i(1-\bar{p}_{t+1})^{1-i}  \bar{q}_{t+1}^j(1-\bar{q}_{t+1})^{1-j} ,
\end{equation}
in which $\{i,j\} \in \{0, 1\}$ too.

Comparing to experimental distributions directly, the theoretical probabilities are multiplied by the observation $\Omega_{i_{0}j_{0}}$ (referring to  Table~\ref{tab:RothPayoffEntropy}) to gain the theoretical distributions.
Fig.~\ref{fig:Rothquivershiyi} (c)  provides an example to illustrate the theoretical distribution of backward transitions using the $\bar{x}_{t-1}$ in Fig.~\ref{fig:Rothquivershiyi} (b) and Eq.~\ref{eq:a4}; Similarly, Fig.~\ref{fig:Rothquivershiyi} (g) illustrates the theoretical distribution of forward transitions using $\bar{x}_{t+1}$ in Fig.~\ref{fig:Rothquivershiyi} (f)  and  Eq.~\ref{eq:a5}; Multiplied factor $\Omega$ refers to the figure in last 4 columns of Table~\ref{tab:RothPayoffEntropy}.

In summary, according to Eq.~\ref{eq:a4} and Eq.~\ref{eq:a5}, together with Tab~\ref{tab: mean start points and terminal points } as constraints, theoretical probabilities of the transitions can be obtained.
Multiplied by the observation $\Omega$ at the given state,  the distribution of the transitions can be obtained and listed in Table~\ref{tab:expected frequencies}.

\begin{table}[htbp2]
\centering
\begin{threeparttable}
\small
\caption{\label{tab:expected frequencies} Expected transitions frequencies of 11 game by MaxEnt}
\begin{tabular}{|c|ccccccccccc|}
  \hline
   \hline
$T_-$&~~g1	&~~g2	&~~g3	&~~g4	&~~g5	&~~g6	&~~g7	&~~g8	&~~g9	&~g10	&~g11		 \\
  \hline
$T_{00{\leftarrow}00}$	&~459	&~754	&~198	&~302	&~116	&~505	&~145	&~106	&~564	 &~138	 &~197	\\
$T_{00{\leftarrow}01}$	&~160	&~~62	&~~59	&~~79	&~~76	&~124	&~167	&~100	&~109	 &~117	 &~240	\\
$T_{00{\leftarrow}10}$	&~279	&~514	&~313	&~205	&~215	&~407	&~~87	&~~75	&~404	 &~101	 &~181	\\
$T_{00{\leftarrow}11}$	&~~97	&~~43	&~~93	&~~54	&~141	&~100	&~100	&~~70	&~~78	 &~~85	 &~219	\\
$T_{01{\leftarrow}00}$	&~~50	&~~84	&~~50	&~248	&~195	&~237	&~119	&~~30	&~255	 &~~34	 &~152	\\
$T_{01{\leftarrow}01}$	&~111	&~~50	&~~74	&1045	&~235	&~300	&1176	&~245	&~373	 &~130	 &~213	\\
$T_{01{\leftarrow}10}$	&~~83	&~~71	&~~85	&~~61	&~209	&~~75	&~~51	&~~43	&~~93	 &~~64	 &~228	\\
$T_{01{\leftarrow}11}$	&~188	&~~43	&~126	&~254	&~250	&~~96	&~497	&~346	&~137	 &~238	 &~322	\\
$T_{10{\leftarrow}00}$	&~415	&~470	&~353	&~~56	&~~90	&~235	&~184	&~179	&~148	 &~195	 &~283	\\
$T_{10{\leftarrow}01}$	&~~69	&~~51	&~129	&~~54	&~104	&~~86	&~~60	&~~72	&~~95	 &~128	 &~261	\\
$T_{10{\leftarrow}10}$	&1007	&1698	&1076	&~247	&~444	&1030	&~439	&~848	&~687	 &~407	 &~189	\\
$T_{10{\leftarrow}11}$	&~166	&~184	&~394	&~234	&~515	&~380	&~142	&~343	&~438	 &~265	 &~175	\\
$T_{11{\leftarrow}00}$	&~~42	&~~60	&~~47	&~~32	&~142	&~146	&~112	&~~72	&~159	 &~~32	 &~224	\\
$T_{11{\leftarrow}01}$	&~121	&~~92	&~~87	&~438	&~481	&~209	&~379	&~212	&~314	 &~134	 &~180	\\
$T_{11{\leftarrow}10}$	&~318	&~123	&~497	&~~80	&~290	&~231	&~190	&~443	&~213	 &~467	 &~290	\\
$T_{11{\leftarrow}11}$	&~926	&~192	&~910	&1102	&~988	&3~33	&~643	&1307	&~424	 &1956	 &~234	\\
  \hline
$T_+$&~~g1	&~~g2	&~~g3	&~~g4	&~~g5	&~~g6	&~~g7	&~~g8	&~~g9	&~g10	&~g11		 \\
  \hline
$T_{00{\rightarrow}00}$	&~452	&~749	&~187	&~309	&~108	&~508	&~142	&~v97	&~563	 &~125	 &~197	\\
$T_{00{\rightarrow}01}$	&~~67	&~101	&~~32	&~244	&~229	&~247	&~105	&~~25	&~306	 &~~34	 &~148	\\
$T_{00{\rightarrow}10}$	&~413	&~461	&~380	&~~50	&~~67	&~256	&~146	&~183	&~187	 &~223	 &~280	\\
$T_{00{\rightarrow}11}$	&~~62	&~~62	&~~65	&~~40	&~144	&~124	&~109	&~~48	&~101	 &~~61	 &~212	\\
$T_{01{\rightarrow}00}$	&~143	&~~51	&~~84	&~~92	&~~67	&~107	&~198	&~~96	&~~83	 &~129	 &~252	\\
$T_{01{\rightarrow}01}$	&~118	&~~49	&~~78	&1029	&~218	&~303	&1162	&~263	&~349	 &~131	 &~205	\\
$T_{01{\rightarrow}10}$	&~~95	&~~76	&~~88	&~~40	&~144	&~~78	&~~70	&~~81	&~~83	 &~101	 &~252	\\
$T_{01{\rightarrow}11}$	&~~77	&~~74	&~~83	&~450	&~462	&~218	&~410	&~223	&~345	 &~104	 &~204	\\
$T_{10{\rightarrow}00}$	&~275	&~531	&~300	&~202	&~238	&~382	&~107	&~~85	&~345	 &~~88	 &~170	\\
$T_{10{\rightarrow}01}$	&~~77	&~~42	&~127	&~~61	&~160	&~~86	&~~48	&~~47	&~102	 &~~90	 &~243	\\
$T_{10{\rightarrow}10}$	&1020	&1695	&1073	&~249	&~452	&1029	&~460	&~843	&~708	 &~405	 &~203	\\
$T_{10{\rightarrow}11}$	&~287	&~133	&~455	&~~76	&~303	&~232	&~210	&~468	&~211	 &~412	 &~291	\\
$T_{11{\rightarrow}00}$	&~~53	&~~45	&~~62	&~~55	&~137	&~114	&~~88	&~~81	&~133	 &~~44	 &~226	\\
$T_{11{\rightarrow}01}$	&~242	&~~53	&~128	&~256	&~280	&~~95	&~492	&~321	&~133	 &~266	 &~311	\\
$T_{11{\rightarrow}10}$	&~201	&~168	&~441	&~234	&~488	&~390	&~113	&~327	&~422	 &~321	 &~165	\\
$T_{11{\rightarrow}11}$	&~909	&~201	&~908	&1104	&~994	&~322	&~631	&1303	&~422	 &1957	 &~229	\\
  \hline
  \hline
\end{tabular}
\begin{tablenotes}
  \item[$\S$] g1 indicates game 1, the rest analogize.
  \end{tablenotes}
\end{threeparttable}
\end{table}

\section{Results}

To test MaxEnt is to evaluate the goodness of fit between the experimental data (in Table~\ref{tab:Actual Frequencies}) and theoretical data (in Table~\ref{tab:expected frequencies}).

\begin{figure}
\centering
\includegraphics[angle=0,width=4.1cm]{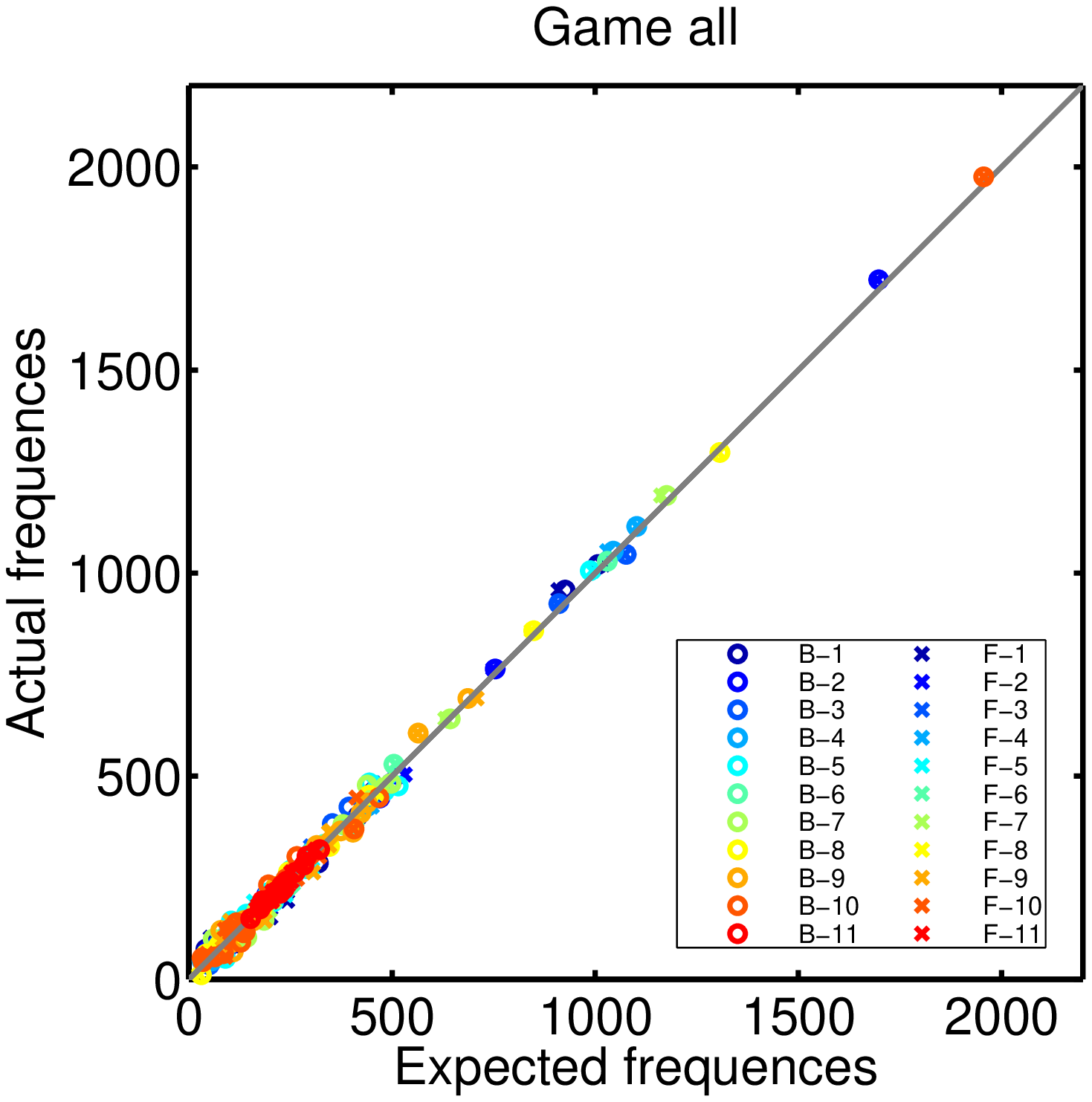}
\includegraphics[angle=0,width=4.1cm]{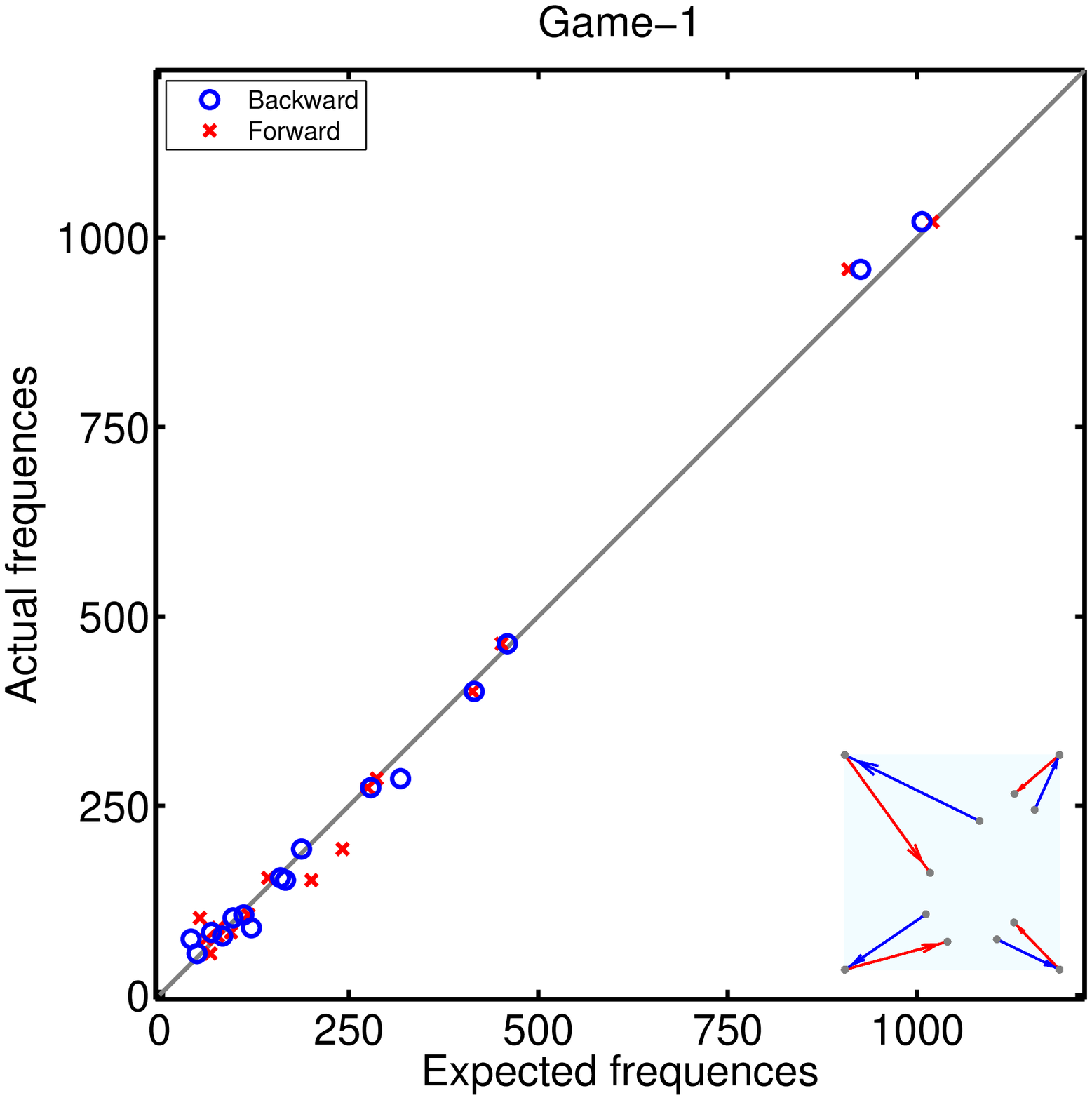}
\includegraphics[angle=0,width=4.1cm]{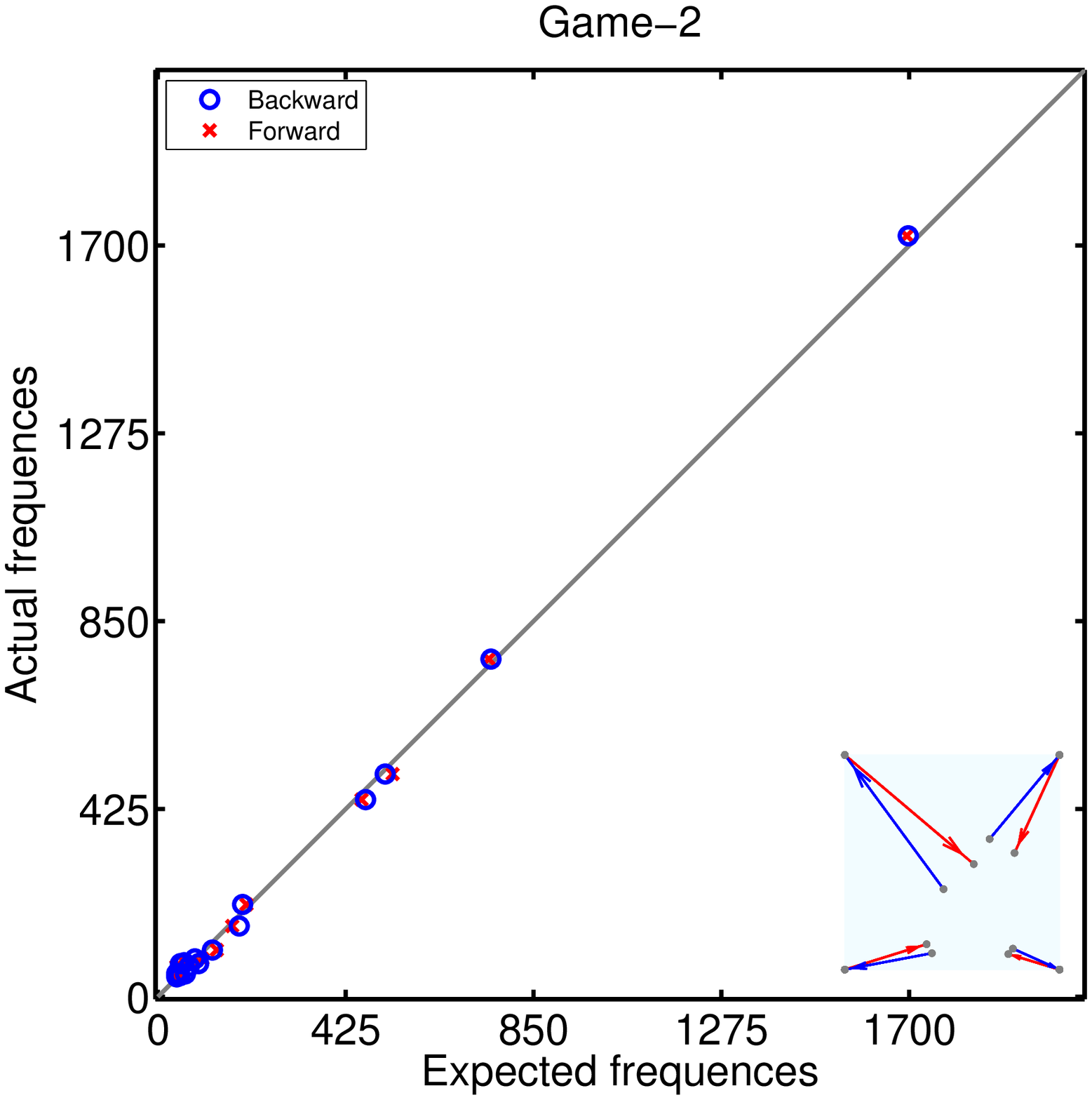}
\includegraphics[angle=0,width=4.1cm]{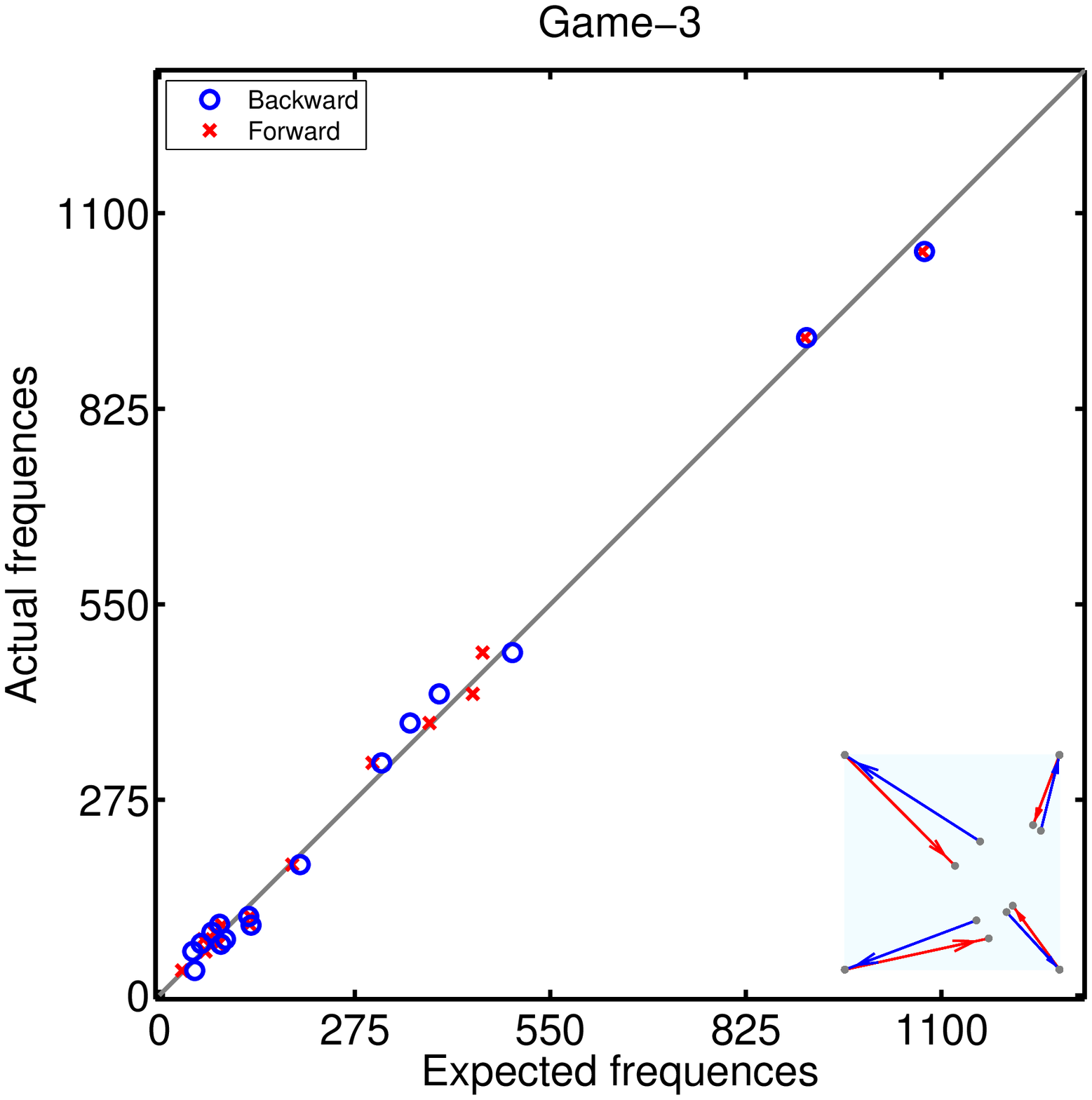}
\includegraphics[angle=0,width=4.1cm]{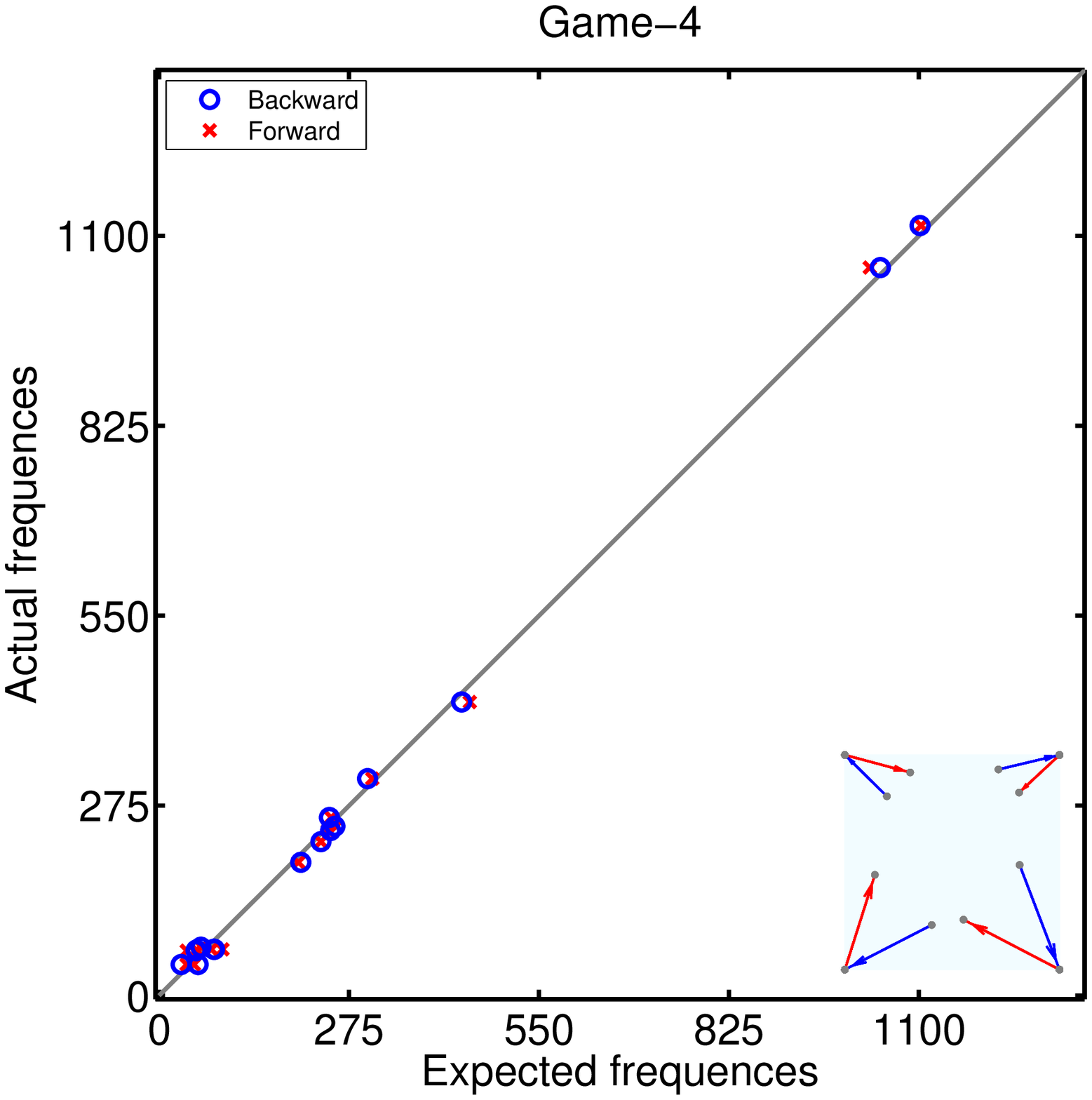}
\includegraphics[angle=0,width=4.1cm]{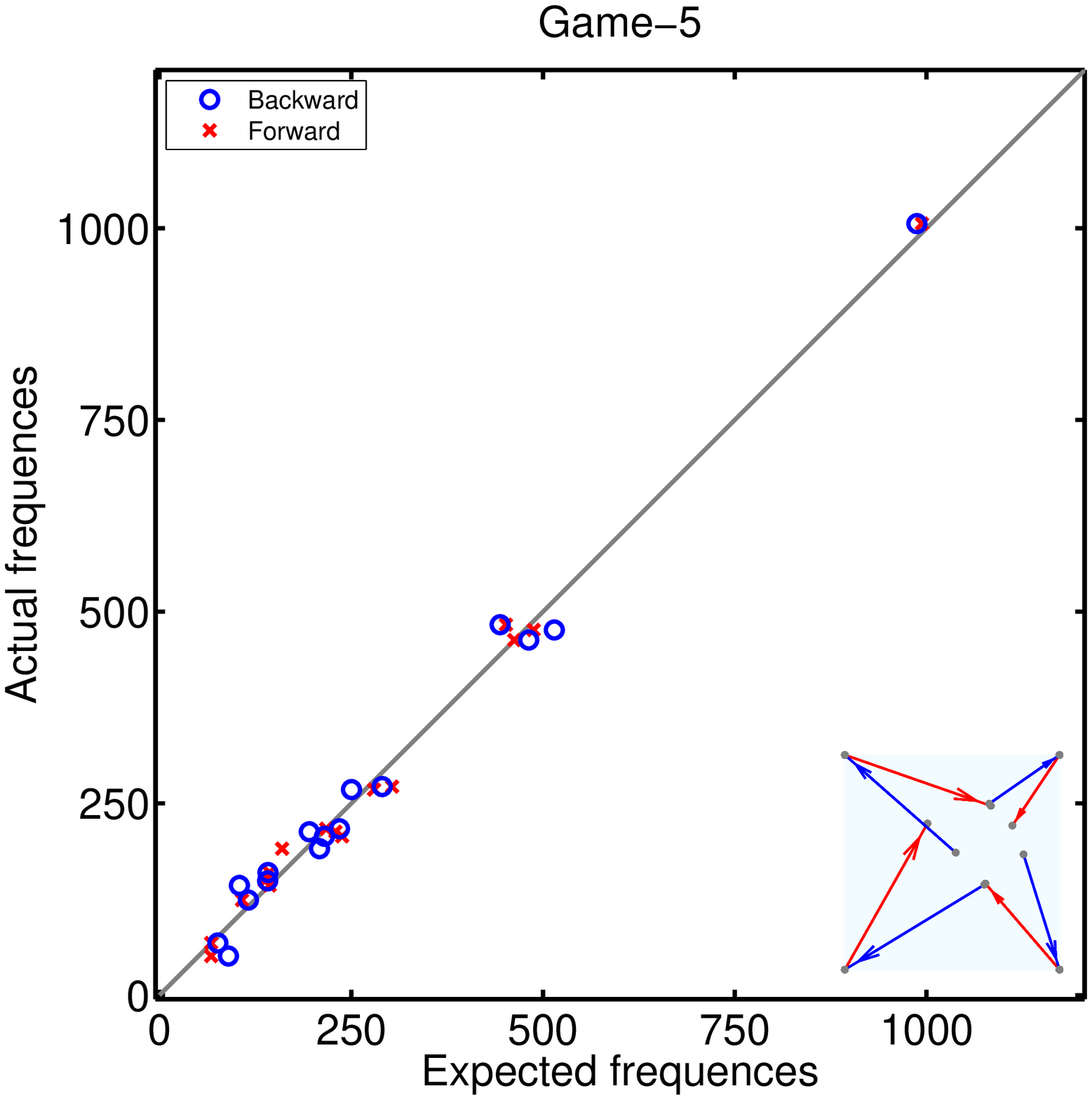}
\includegraphics[angle=0,width=4.1cm]{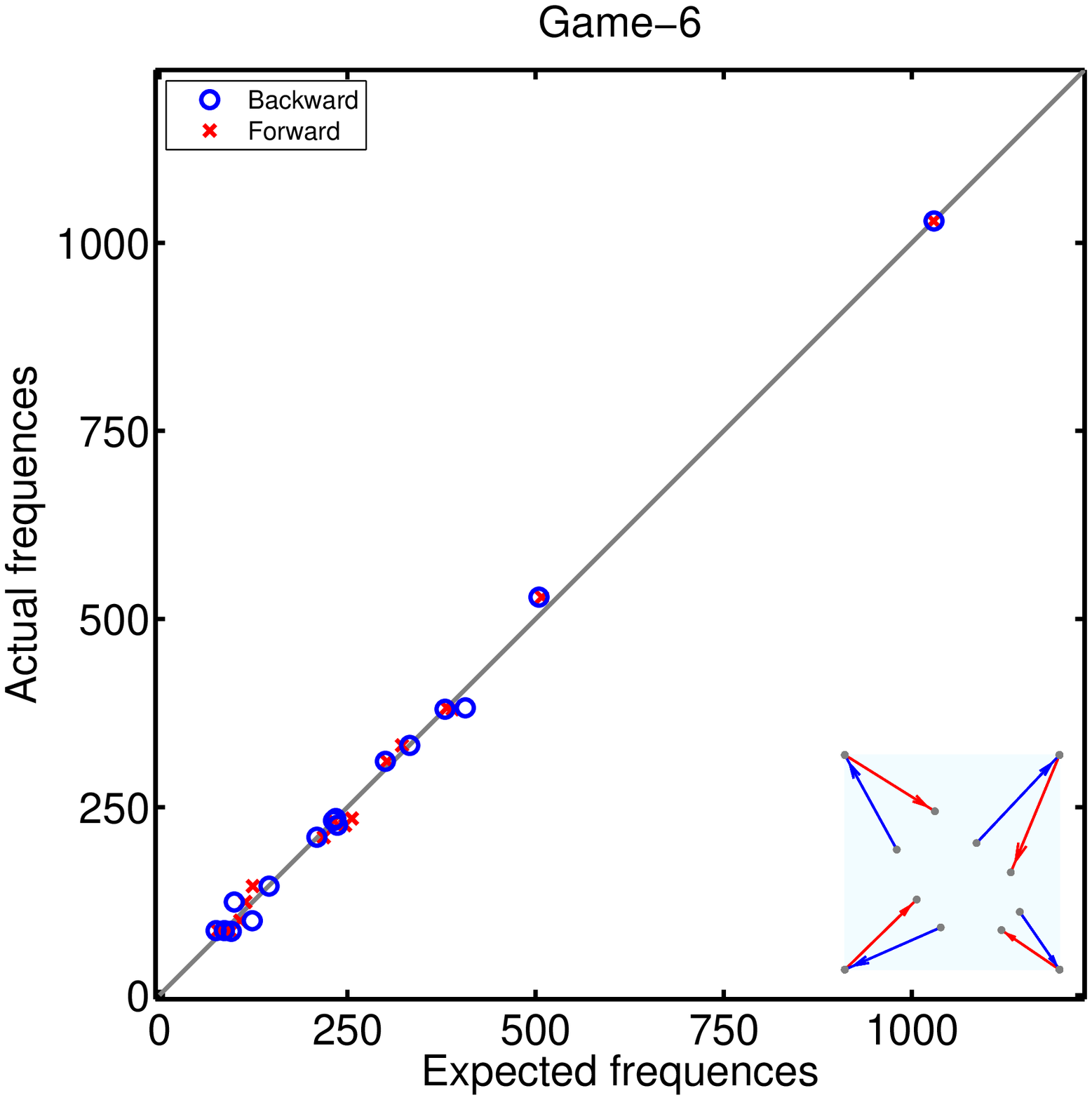}
\includegraphics[angle=0,width=4.1cm]{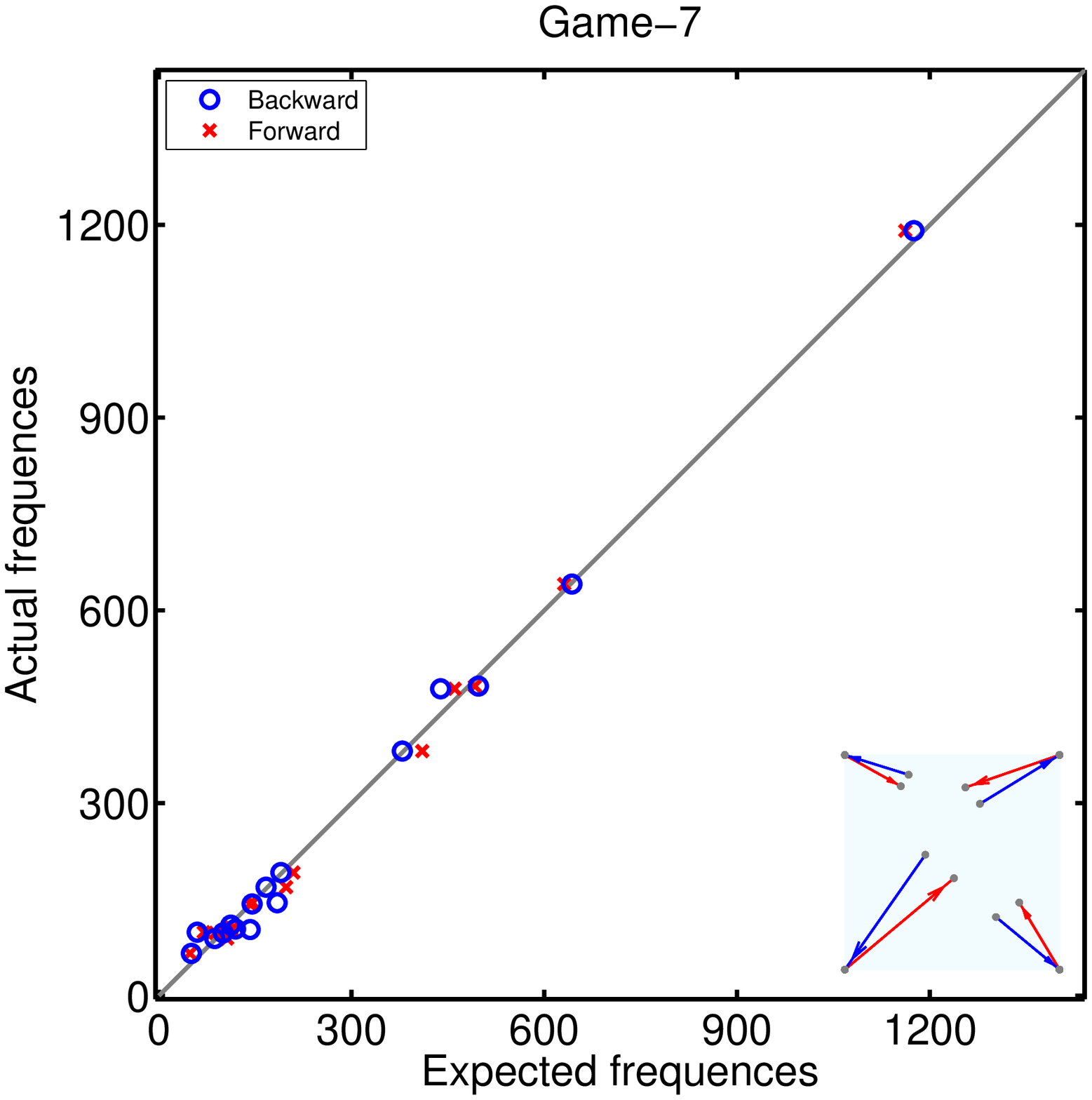}
\includegraphics[angle=0,width=4.1cm]{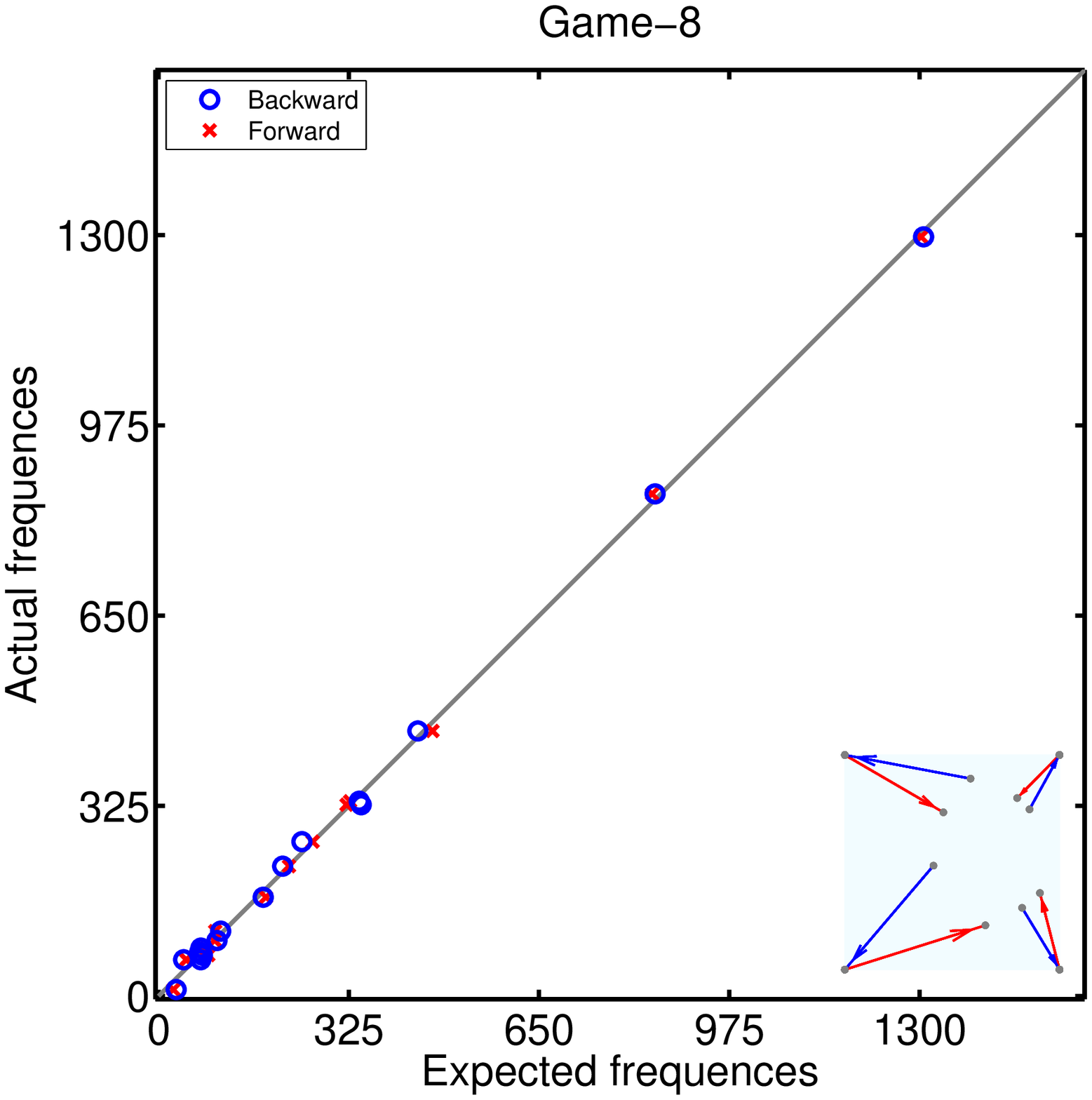}
\includegraphics[angle=0,width=4.1cm]{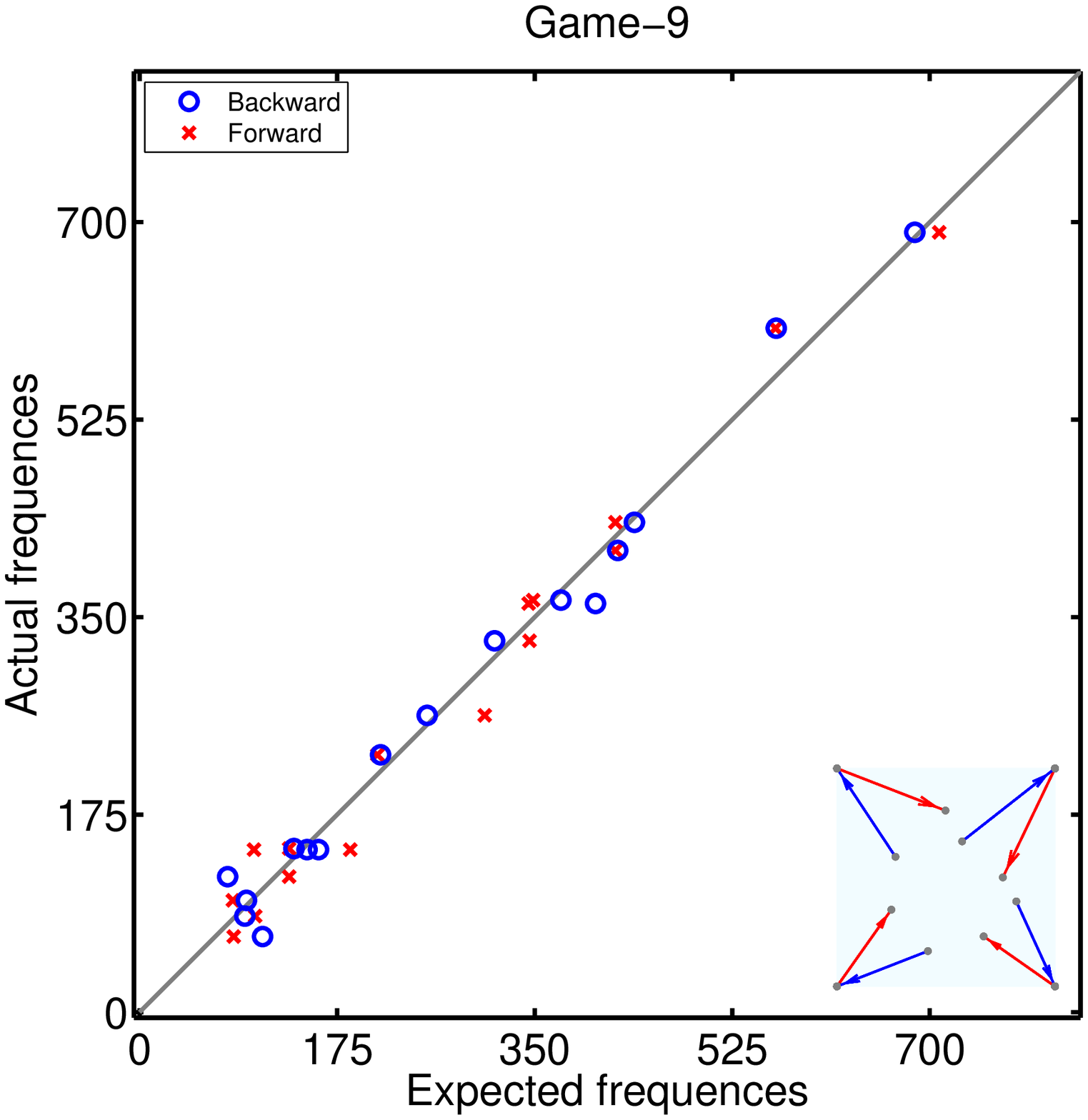}
\includegraphics[angle=0,width=4.1cm]{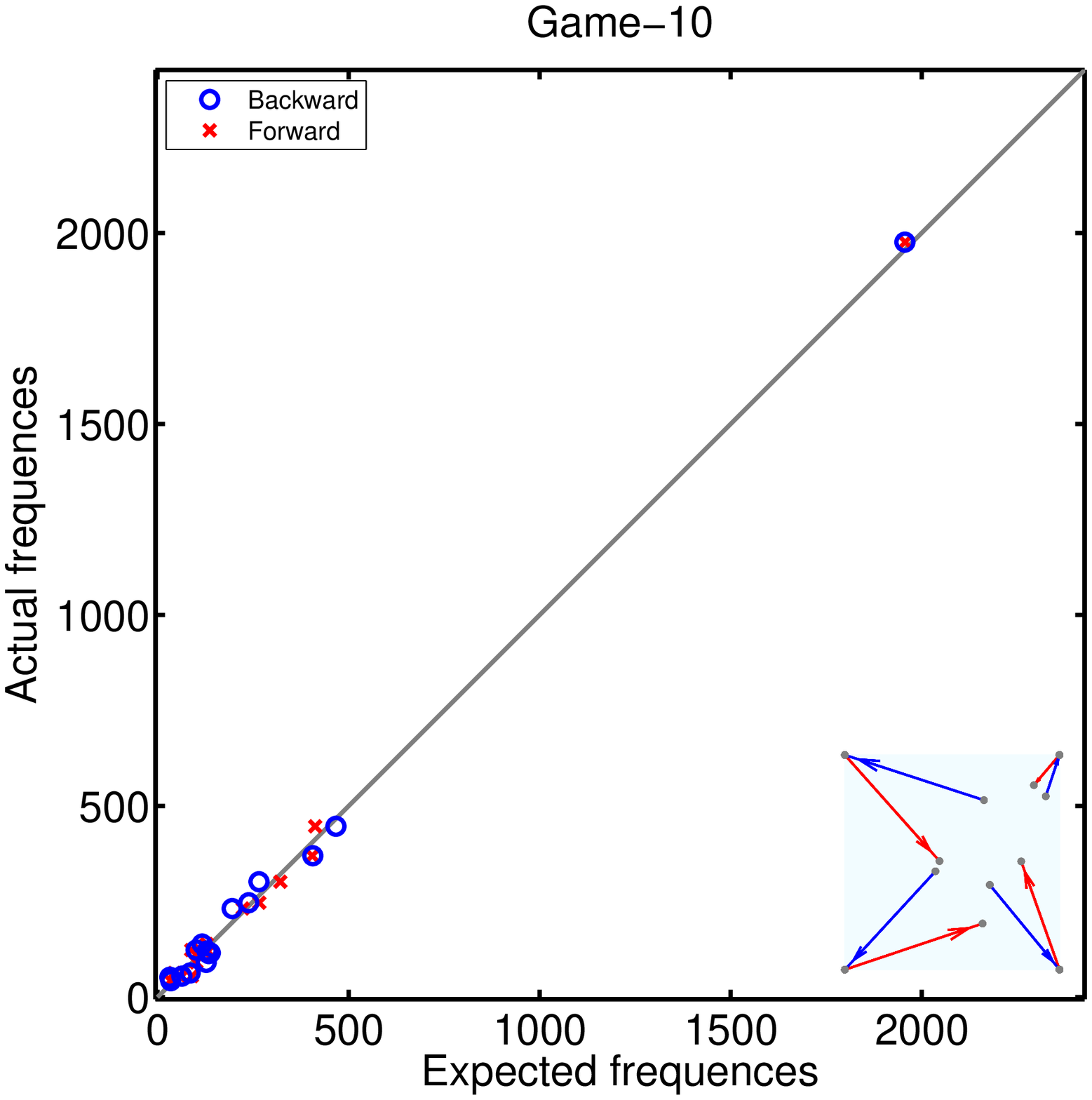}
\includegraphics[angle=0,width=4.1cm]{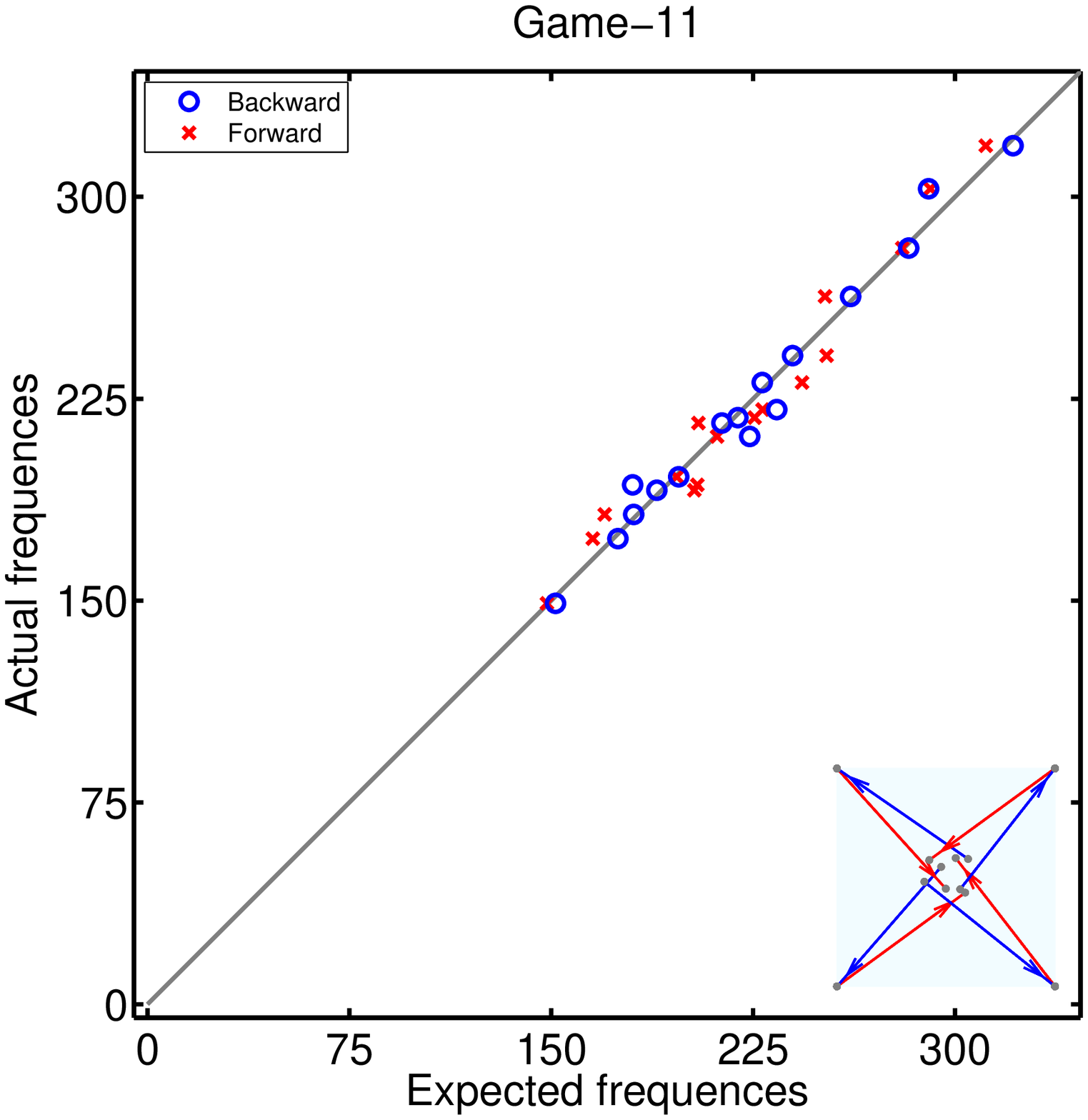}
\caption{Comparison of the actual transition frequencies with expected transition frequencies in the experimental games. Horizonal axis is the expected transition frequencies from MaxEnt hypothesis; meanwhile, vertical axis is the actual transition frequencies. The first figure is the results of all 11 games. Figure 2$nd$ to last is for game 1 to game 11, respectively. In the first figure, symbol B (F) in the legend indicates backward (forward). For each game, there are four social states and 16 kinds of social forward transitions and 16 kinds of backward transitions. The cycles in blue (cross in red) indicate the backward transitions (forward transitions).
A dot in diagonal line means the experimental transition frequencies equal to the theoretical transition frequencies.
The sub-figures in the right down corners are the schematic diagrams for aggregated backward (forward) transitions  in blue arrows (in red arrows).}
\label{fig:roth}
\end{figure}

Fig.~\ref{fig:roth} plots the results of observed experimental transition frequencies (in horizon, $x$-axis) and theoretical transition frequencies(vertical, $y$-axis).
The first figure is the results for all 11 games, and from second to last is game 1 to game 11, respectively.
For each game, there are 32 samples of social strategy transitions.
The cycles in blue indicate the backward transitions and the crosses in red indicate the forward transitions. 
Significantly, all of the backward transition samples (blue cycles) and forward transitions samples (red crosses) are close to the diagonal lines which means theoretical values from MaxEnt are close to experimental values.

The liner regression results are shown in Table\ref{tab:Results of linear regression for $T_{f}$ of each game}. Obliviously, each of the liner regression coefficients is very close to 1 and the $p<0.001$. 
Table\ref{tab:Results of linear regression for $T_{f}$ of each game} also provides the  $99\%$ C.I. (confidence interval) both for liner regression coefficients and intercept constant. All the lower bound of $99\%$ C.I. of regression coefficients are smaller than but very close to 1 and the upper bound are larger than but also very close to 1; then the equal hypothesis of the two variables can not be rejected. Meanwhile, for intercept constant (y-intercept, the point where a line crosses the y-axis), none of the $p$ value is smaller than 0.42, all of the lower bound of the $99\%$  C.I. are smaller than 0 and all of the upper bound are larger than 0. Then, the hypothesis that the regression line cross $0$ can not be rejected. These statistical results indicate that the hypothesis that theoretical values statistically equal to experimental observation of the transitions is supported.

In summary, in the laboratory experimental two-person constant sum 2$ \times$2 games, the outcome of the social transitions fit MaxEnt. In another word, given the mean vector of the transitions of a given state, the distributions of the all transitions of the given state can be estimated with MaxEnt, and fit experiments data exactly.

\begin{table}[htbp2]
\centering
\begin{threeparttable}
\small
\caption{\label{tab:Results of linear regression for $T_{f}$ of each game} Results of linear regression for $T_-$ and  $T_+$ of each game}
\begin{tabular}{|c|ccc|ccc|ccccccccccc}
  \hline
   \hline
&Coef. $T_-$\dag	&[99\%C.I]\S	&const.~[99\%C.I]\ddag	&Coef. $T_+$\dag	&[99\%C.I]\S	 &const.~[99\%C.I]\ddag\\
   \hline
~g1&1.019	&0.971~~~1.067	&-24.497~~~13.829	&1.020	&0.950	~~~1.090	&-33.665	~~~22.369	 \\
~g2&1.010	&0.982~~~1.039	&-17.453~~~11.596	&1.012	&0.983	~~~1.041	&-17.993	~~~11.337	 \\
~g3&0.995	&0.943~~~1.047	&-19.863~~~22.717	&0.997	&0.952	~~~1.041	&-17.420	~~~19.273	 \\
~g4&1.009	&0.981~~~1.037	&-14.445~~~~9.388	&1.013	&0.978	~~~1.048	&-18.533	~~~11.146	 \\
~g5&1.005	&0.922~~~1.088	&-31.300~~~28.440	&1.007	&0.942	~~~1.072	&-25.466	~~~21.471	 \\
~g6&1.002	&0.956~~~1.049	&-17.457~~~16.123	&1.003	&0.961	~~~1.045	&-16.216	~~~14.447	 \\
~g7&1.012	&0.954~~~1.070	&-26.656~~~19.858	&1.017	&0.970	~~~1.065	&-23.895	~~~14.080	 \\
~g8&0.997	&0.968~~~1.026	&-11.762~~~13.329	&1.000	&0.974	~~~1.026	&-11.121	~~~11.139	 \\
~g9&1.009	&0.909~~~1.110	&-36.096~~~30.798	&1.006	&0.894	~~~1.118	&-38.853	~~~35.654	 \\
g10&1.008	&0.966~~~1.050	&-24.502~~~20.038	&1.009	&0.972	~~~1.045	&-21.852	~~~16.953	 \\
g11&1.002	&0.886~~~1.119	&-27.159~~~26.109	&1.011	&0.848	~~~1.175	&-39.890	~~~34.780	 \\
    \hline
total&1.007	&0.995~~~1.019	&-6.700~~~~2.889	&1.009	&0.997  ~~~1.020	&-7.170     ~~~~2.317	 \\
    \hline
 \hline
  \end{tabular}
  \begin{tablenotes}
  \item[\dag] liner regression coefficient.
   \item[\S]  99$\%$ Confident Interval  for liner regression coefficient.
      \item[\ddag] 99$\%$ Confident Interval for intercept constant.
  \end{tablenotes}
     \end{threeparttable}
  \end{table}

\section{Discussion and conclusion}

The main result of this report is that, the  social strategy transitions
are not erratic but governed by MaxEnt suggested by Jaynes~\cite{Jaynes1957}. This result comes from the dynamical observables in the experiments of human subject competing games~\cite{RothErev2007,XuetalMaxEnt2012}. 

\subsection{Examples of the Necessary of MaxEnt}
To make the MaxEnt in social dynamics easier to understand, we provides two alternative examples in Fig.~\ref{fig:Rothquivershiyi}; The first example is in sub-figure (d), the backward transition distribution, which is consistent with the aggregated backward transition shown in (b) but not capture the experiment result in (a);
 In another word, in (d), the frequencies for $T_{00\leftarrow01}$, $T_{00\leftarrow00}$, $T_{00\leftarrow10}$   and $T_{00\leftarrow11}$ could be 133, 28, 0 and 271; Even though this distribution satisfied the constraint condition (b), it is far away from the experimental distribution in (a).  Second example is for forward state transition in (h); without MaxEnt, a plausible distribution like the (h) in Fig.~\ref{fig:Rothquivershiyi} does not provide efficient information of experimental dynamical observable in (e) with the constraint condition of (f).  Alternatively, with MaxEnt, results in (c) and (g), by using (b) and (f) as the constraints, can recover the experimental distribution in (a) and (e), respectively.
%
%

%
%
%

\subsection{MaxEnt in social  state transition and experimental social dynamics}

In this section, we explain the connection of present results to the results found in experimental social dynamics.

%
The social dynamics of human subjects systems is an interdisciplinary field~\cite{Friedman1998Rev,social2009dynamics,Sandholm2011,young2008stochastic}.
In this field, evolutionary game theory provides a general mathematical framework for the theoretical investigation of social dynamics and is used commonly by physicists and economics.
However, this theory has rare gain the supports from laboratory experiments\footnote{One point need to emphasis, most of the experiments are conducted by the social scientists in the field called as \emph{experimental economics}. All the experiments are the incentivized laboratory experiment using human subjects. Traditional experimental testing on social dynamics mainly focus on the convergence property of the equilibrium (e.g.,~\cite{VSmith1989,yan2004,CT98,RothSpeed2006}). Early experiments had demonstrated the qualitative consistence between the evolutionary dynamics and laboratory social behaviors~\cite{Friedman1998,Friedman1996,Huyck2008,HuyckSamuelson2001} but not the quantitative  consistence. In  experiment data, as pointed out~\cite{benaim2009learning}, it is difficult to test out the dynamics patterns (e.g., cycles~\cite{benaim2009learning,Friedman2012}) which are  predicted by evolutionary game theory.} of human subject social systems quantitatively.

Only quite recently, according to the three reports from three independent research groups, quantitative experimental testing on evolutionary dynamics is becoming possible. The three reports are following. (1) The first is the report from Hoffman, Suetens, Nowak and Gneezy (2012)~\cite{Nowak2012}. In three   Rock-paper-scissors games, 
the authors compare behavior with three different symmetric matrices whose mean distances from identical Nash equilibrium (NE) are equal (unequal) in classical (evolutionary) game theory. They find the mean distance from NE in a treatment is larger which is predicted by replicator dynamics model in evolutionary game theory. This is the first experimental report to support one of the most fundamental concept --- Evolutionarily stable state (ESS) --- in evolutionary game theory.  In their report, the simplest replicator dynamics model is used as a reference. (2) The second is the report from Cason, Friedman and Hopkins (2012)~\cite{Friedman2012}. Using continuous time experiments, also in Rock-Paper-Scissors games, the authors found cycles directly. More Importantly, they found that the cycle amplitude, frequency and direction are consistent with standard learning models\footnote{This findings of cycles in Rock-Paper-Scissors games~\cite{Friedman2012} are supported by the discrete time experiments of three different parameters Rock-Paper-Scissors games~\cite{XuWangCycleRPS,XuWangAsymmetryRPS} from an independent research group.}. In their report~\cite{Friedman2012}, the logit dynamics model is used as a reference. Another important point is "time" are served as controlled variable in their experiments.  (3) The third are the report from Xu, Wang and Wang (2012)~\cite{XuWangWang1208}. In two-population random-matched 2$\times$2 games with 12 different payoff-matrix parameters~\cite{selten2008}, the experimental frequencies is found to be linear positive related to the theoretical frequencies significant ($>$5$\sigma$). The payoff-normalized replicator dynamics model is used as the reference. Together with the observed cyclic velocity vector field pattern in experiments~\cite{XuWang2011ICCS}, evidences from 2$\times$2 games support the evolutionary game theory as well. To test evolutionary dynamics in laboratory human subject social systems,  these are the three experiments which are reported recently.

Notice that, all the three groups use accumulated observable (macro observable) to describe the social dynamics behaviors. In this letter,  the macro observable which are used as the constraint conditions (e.g., the mean  aggregated forward transition) are also macro observable. In this letter, we show that MaxEnt  can provides more dynamics information (micro observable, the state-to-state transits) from the limited accumulated observable (macro observable) in the experiments.

In words, present report might provide a paradigm --- the micro and macro dynamical observable could be linked by MaxEnt in social dynamical processes in experiments. We suggest the results reported in this letter could be replicated in more general conditions in the experiments of human subject social dynamics.




%
%
\subsection{MaxEnt as a Link  between Nature and  Social Science}
In economics, MaxEnt approach has gained its widely applications, e.g., in  market equilibrium~\cite{Toda2010,Barde2012}, in wealth and income distribution~\cite{Castaldi2007,WuMaxEntIncome2003}, in firm growth rates~\cite{Alfarano2008} in behavior modeling~\cite{Wolpert2012}. Theoretical interpreting or modeling of the distributions of social outcomes with MaxEnt is growing.

 However, to the best of our knowledge, the dynamic behavior (both of the backward and forward transitions) obeys MaxEnt --- this point has never been empirical presented. Our finding of the MaxEnt in dynamic behaviors in experimental data can be an encourage information for investigating the potential self-consistence of social outcome --- both in static and dynamic performance.
%

%

MaxEnt, as a technique, can be used to predict the geographic distribution of
any spatial phenomena, including plants and animals~\cite{MaxEntSpecies2006}. In game theory condition, the spatial phenomena of social behavior is the phenomena in strategy space, at the same time, the absence or appearance of strategy is relative to  the absence or appearance of species. This picture has been well built~\cite{Sandholm2011} to unify the evolutionary theorems in biology science and social science. Our findings of the social behavior fitting MaxEnt, both in dynamic respect in this report and in static respect in~\cite{XuetalMaxEnt2012}, suggest that human subject social systems and natural systems could have wider common backgrounds.

For the future investigations, several points need to be considered. As we have shown in static~\cite{XuetalMaxEnt2012} and in one step ($x_{t\pm 1}$) dynamics social behaviors obey MaxEnt, dose any step transitions always obey MaxEnt? What is the bound of the MaxEnt in social interaction systems?

One can notice that, in the 11 games, all the social environments are different (for the payoff matrix are different), all the mixed strategy Nash equilibrium are different, however, all the social transitions obeying MexEnt are indifferent.

\subsection{Conclusion}
By employing experimental economics data, we test the MaxEnt hypothesis in social transitions. In the experimental constant sum two-person 2$\times$2 games, the results show that, not only static social state distributions obey MaxEnt~\cite{XuetalMaxEnt2012},  the distributions of the social state transitions also fit MaxEnt. This  finding suggests that MaxEnt can also be an approach for the social dynamics.\\

\textbf{Acknowledgment:} Thanks to Alvin Roth for kindly providing us the data. This research was supported by the grant from the Experimental Social Science Laboratory of Zhejiang University (2012 annual project: The property of unstable equilibrium: an experimental investigation). We thank the anonymity reviewer for helpful suggestion. We also thank Zunfeng Wang for technical assistance, thank Shuang Wang and Yuqing Luo for polishing the English.  All remaining errors belong to the authors.

\bibliographystyle{plain}


%

\end{document}